\def \spc {\;\;}

\def \Pt {$p_{T}$ }

\def \sumet {$\sum E_{T}$ }
\def \misset {$\not\!\!E_{T}$ }
\documentstyle[doublespace,12pt,epsf]{article}
\begin{document}
\begin{flushright}
  \baselineskip 0.5cm
   FERMILAB-PUB-96/098-E\\
   MAY 8, 1996
\end{flushright}
\begin{center}
\vspace{1cm}
\large{\bf{Further Properties of High-Mass Multijet Events}}\\
\large{\bf{ at the Fermilab Proton-Antiproton Collider}}
\end{center}
\baselineskip 0.8cm
\font\eightit=cmti8
\def\r#1{\ignorespaces $^{#1}$}
\hfilneg
\begin{sloppypar}
\noindent
F.~Abe,\r {14} H.~Akimoto,\r {32}
A.~Akopian,\r {27} M.~G.~Albrow,\r 7 S.~R.~Amendolia,\r {23} 
D.~Amidei,\r {17} J.~Antos,\r {29} C.~Anway-Wiese,\r 4 S.~Aota,\r {32}
G.~Apollinari,\r {27}
% ...
\end{sloppypar}

\vskip .025in
\begin{center}
(CDF Collaboration)
\end{center}

\vskip .025in
\begin{center}
\r 1  {\eightit Argonne National Laboratory, Argonne, Illinois 60439} \\
\r 2  {\eightit Istituto Nazionale di Fisica Nucleare, University of Bologna,
I-40126 Bologna, Italy} \\
\r 3  {\eightit Brandeis University, Waltham, Massachusetts 02254} \\
\r 4  {\eightit University of California at Los Angeles, Los 
Angeles, California  90024} \\  
\r 5  {\eightit University of Chicago, Chicago, Illinois 60637} \\
\r 6  {\eightit Duke University, Durham, North Carolina  27708} \\
\r 7  {\eightit Fermi National Accelerator Laboratory, Batavia, Illinois 
60510} \\
\r 8  {\eightit Laboratori Nazionali di Frascati, Istituto Nazionale di Fisica
               Nucleare, I-00044 Frascati, Italy} \\
\r 9  {\eightit Harvard University, Cambridge, Massachusetts 02138} \\
\r {10} {\eightit Hiroshima University, Higashi-Hiroshima 724, Japan} \\
\r {11} {\eightit University of Illinois, Urbana, Illinois 61801} \\
\r {12} {\eightit Institute of Particle Physics, McGill University, Montreal 
H3A 2T8, and University of Toronto,\\ Toronto M5S 1A7, Canada} \\
\r {13} {\eightit The Johns Hopkins University, Baltimore, Maryland 21218} \\
\r {14} {\eightit National Laboratory for High Energy Physics (KEK), Tsukuba, 
Ibaraki 305, Japan} \\
\r {15} {\eightit Ernest Orland Lawrence Berkeley Laboratory, Berkeley, 
        California 94720} \\
\r {16} {\eightit Massachusetts Institute of Technology, Cambridge,
Massachusetts  02139} \\   
\r {17} {\eightit University of Michigan, Ann Arbor, Michigan 48109} \\
\r {18} {\eightit Michigan State University, East Lansing, Michigan  48824} \\
\r {19} {\eightit University of New Mexico, Albuquerque, New Mexico 87131} \\
\r {20} {\eightit Osaka City University, Osaka 588, Japan} \\
\r {21} {\eightit Universita di Padova, Istituto Nazionale di Fisica 
          Nucleare, Sezione di Padova, I-35131 Padova, Italy} \\
\r {22} {\eightit University of Pennsylvania, Philadelphia, 
        Pennsylvania 19104} \\   
\r {23} {\eightit Istituto Nazionale di Fisica Nucleare, University and Scuola
               Normale Superiore of Pisa, I-56100 Pisa, Italy} \\
\r {24} {\eightit University of Pittsburgh, Pittsburgh, Pennsylvania 15260} \\
\r {25} {\eightit Purdue University, West Lafayette, Indiana 47907} \\
\r {26} {\eightit University of Rochester, Rochester, New York 14627} \\
\r {27} {\eightit Rockefeller University, New York, New York 10021} \\
\r {28} {\eightit Rutgers University, Piscataway, New Jersey 08854} \\
\r {29} {\eightit Academia Sinica, Taipei, Taiwan 11529, Republic of China} \\
\r {30} {\eightit Texas A\&M University, College Station, Texas 77843} \\
\r {31} {\eightit Texas Tech University, Lubbock, Texas 79409} \\
\r {32} {\eightit University of Tsukuba, Tsukuba, Ibaraki 305, Japan} \\
\r {33} {\eightit Tufts University, Medford, Massachusetts 02155} \\
\r {34} {\eightit University of Wisconsin, Madison, Wisconsin 53706} \\
\r {35} {\eightit Yale University, New Haven, Connecticut 06511} \\
\end{center}

\vspace{0.4cm}
PACS numbers: 12.38Qk, 13.85.-t, 13.85.Hd, 13.87.-a
\large
\normalsize
\vspace{0.6cm}
\begin{center}
  \large{Abstract}
\end{center}
  \vspace{0.5cm}
  The properties of high-mass multijet events produced at the 
  Fermilab proton-antiproton collider are compared with 
  leading order QCD matrix element predictions, 
  QCD parton shower 
  Monte Carlo predictions, and the predictions from a model in which 
  events are distributed uniformly over the available multibody 
  phase-space. 
  Multijet distributions corresponding to (4N-4) 
  variables that span the N-body parameter space are found to be 
  well described by the 
  QCD calculations for 
  inclusive three-jet, four-jet, and five-jet events. 
  The agreement between data, QCD Matrix Element calculations, and QCD 
  parton shower Monte Carlo predictions suggests that $2 \rightarrow 2$ 
  scattering plus gluon radiation provides a good first approximation to 
  the full LO QCD matrix element for events with three, four, or even five 
  jets in the final state. 
\newpage

\begin{section} {Introduction}
A study of the properties of events containing three-or-more jets 
produced in high-energy hadron-hadron collisions can 
(i) test our understanding of the higher-order QCD processes that result 
in multijet production, 
(ii) test the QCD parton shower Monte Carlo approximation to the full 
leading order (LO) QCD matrix element, and 
(iii) enable a search for new phenomena associated with the presence 
of many hard partons in the final state. 
The CDF collaboration has recently reported the characteristics  
of two-jet, three-jet, four-jet, five-jet, and six-jet events 
\cite{multijet_paper} produced at the Tevatron proton-antiproton collider 
operating at a center-of-mass energy of 1.8 TeV. Results from an analysis of 
events with multijet masses exceeding 600 GeV/$c^2$ were presented for a 
data sample corresponding to an integrated luminosity of 35 pb$^{-1}$. 
The multijet-mass distributions, leading-jet angular distributions, and mass 
dependent jet multiplicity distributions 
were shown to be well described by both the NJETS~\cite{NJETS} 
LO QCD matrix element calculation for events with up to five jets, and 
the HERWIG~\cite{HERWIG} QCD parton shower Monte Carlo calculation 
for events with up to six jets. For these selected distributions the QCD 
predictions were found to give a good description of the data. 

In the present paper we use a larger data sample and 
a more comprehensive set of multijet distributions to 
extend our comparison of the properties of high-mass 
multijet events with QCD predictions. 
In particular, we use the set of (4N-4) variables that span the N-jet
parameter space and were recently proposed by Geer and Asakawa~\cite{preprint}, 
and compare 
the observed three-jet, four-jet, and five-jet event characteristics 
with (a) NJETS LO QCD matrix element predictions, (b) HERWIG parton shower 
Monte Carlo predictions, and (c) predictions from a model in which events 
are uniformly distributed over the available multijet phase-space. 
Results are based on a data sample which was recorded 
by the CDF collaboration during the period 1992 - 1995, and 
corresponds to an integrated luminosity of 105~pb$^{-1}$.

\end{section}

\begin{section} {Experimental Details}

A description of the CDF detector can be found in Ref.~\cite{CDF}.
Full details of the CDF jet algorithm, jet corrections, and jet resolution 
functions can be found in Ref.~\cite{cdf_3jet}, 
and a description of the trigger and event selection requirements for 
the high-mass multijet sample are given in Ref.~\cite{multijet_paper}. 
In the following we give a summary of the main details of the CDF detector, 
jet reconstruction, and event selection requirements that are relevant to 
results presented in this paper. We use the CDF co-ordinate system in 
which the origin is at the center of the detector, the z-axis is along 
the beam direction, $\theta$ is the polar 
angle with respect to the z-axis, and $\phi$ is the azimuthal angle 
measured around the beam direction. 

The multijet analysis described in the following sections exploits the 
CDF calorimeters, which cover the pseudorapidity region $|\eta| < 4.2$, 
where $\eta \equiv -\ln (\tan \theta/2)$. The calorimeters are constructed 
in a tower geometry in $\eta$ - $\phi$ space, and 
are segmented in depth into electromagnetic and hadronic compartments. 
The calorimeter towers 
are 0.1 units wide in $\eta$. The tower widths in $\phi$ are $15^{o}$ in 
the central region and $5^{o}$ at larger $|\eta|$ (approximately 
$|\eta| > 1.2$). 
Jets are reconstructed using an algorithm that forms 
clusters from localized energy depositions in the calorimeter towers. 
Calorimeter towers are associated with a jet if their separation 
from the jet axis in ($\eta,\phi$)-space 
$\Delta$R = $(\Delta\eta^2 + \Delta\phi^2)^{1/2} < R_0$. 
For the analysis described in this paper the clustering cone radius 
$R_0$ = 0.7 was chosen. 
With this $R_0$ a plot of the separation between all jets 
observed in the data sample described below reveals that, to a good 
approximation, clusters with separations 
$\Delta R < 0.8$ are always merged by the jet algorithm into a single jet, 
and clusters with separations  $\Delta R > 1.0$ are never merged.
Thus, the effective minimum observable separation between jets 
$\Delta R_{MIN} = 0.9 \pm 0.1$.
Jet energies are corrected for calorimeter non-linearities, energy lost
in uninstrumented regions and outside of the clustering cone, and 
energy gained from the underlying event. The jet corrections typically 
increase jet energies by $25\%$ for jets with transverse energy 
$E_T  = E \sin\theta >$ 60 GeV, where $\theta$ is the 
angle between the jet axis and the beam direction. The jet corrections 
are larger for lower $E_T$ jets, and typically increase jet energies 
by about $30\%$ ($40\%$) for jets with $E_T =$ 40 GeV (20 GeV). 
After correction, jet energies 
are measured with a precision $\sigma_{E}/E$ of approximately 0.1 and 
multijet masses calculated from the jet four-vectors are measured with a 
precision $\sigma_{m}/m$ of approximately 0.1. 
The systematic uncertainty on the jet energy scale is $5\%$ for jets 
in the central region ($|\eta| < 1.2$). There is an additional systematic 
uncertainty of 2\% on the energy scale of jets at larger $|\eta|$ relative 
to the corresponding scale for central jets.

The data were recorded using a trigger which required 
\sumet $>$ 300 GeV, where the sum is over the transverse 
energies ($E_T$) of all uncorrected jets with 
$E_T >$ 10 GeV, and the jet transverse energies were calculated  
assuming an event vertex at the center of the detector (x=y=z=0).
In the subsequent analysis 
the \sumet was recalculated using the reconstructed vertex 
position and corrected jet energies, and summing over 
all jets with corrected $E_T >$ 20 GeV. 
Events were retained with \sumet $>$ 420 GeV. To reject 
backgrounds from cosmic ray interactions, beam halo, and detector malfunctions, 
the events were required to have 
(i) total energy less than 2000 GeV, 
(ii) a primary vertex reconstructed with $|z| < 60$~cm, 
(iii) no significant energy deposited in the hadron calorimeters 
out-of-time with the proton-antiproton collision, and 
(iv) missing-$E_T$ significance \cite{SUMET_PRD} 
$S \equiv$ \misset/(\sumet)$^{1/2} < 6$, where \misset 
$\equiv |\sum\overline{E_T}_i|$, 
and $\overline{E_T}_i$ is a vector that points from the 
interaction vertex to calorimeter cell i and has magnitude 
equal to the cell $E_T$. These requirements select 
30245  events. 

\end{section}

\begin{section} {QCD and Phase-Space Predictions}

In the following we will compare observed multijet distributions with 
predictions 
from (a) the HERWIG {\cite{HERWIG}} QCD parton shower Monte Carlo program, 
(b) the NJETS {\cite{NJETS}} LO QCD 
$2 \rightarrow N$ matrix element Monte Carlo program, and 
(c) a model in which events are distributed uniformly over the available 
N-body phase-space. 

\begin{subsection} {The HERWIG Parton Shower Monte Carlo Calculation}

HERWIG {\cite{HERWIG}} is a QCD
parton shower Monte Carlo program that includes both
initial- and final-state gluon radiation. 
HERWIG predictions can be thought of as QCD 
$2 \rightarrow 2$ predictions with gluon radiation, color coherence,
hadronization, and an underlying event. 
We have used version 5.6 of the HERWIG Monte Carlo 
program together with a full simulation of the CDF detector response. 
In our HERWIG calculations we have used the CTEQ1M~\cite{cteq}
structure functions and the scale Q$^2$ = stu/2(s$^2$+u$^2$+t$^2$).
HERWIG generates $2 \rightarrow 2$
processes above a specified $p_T^{hard}$ where $p_T^{hard}$ is the \Pt of
the outgoing partons from the hard scatter before any radiation has occurred.
It is important to chose a low value of $p_T^{hard}$ so that adequate 
account is taken of events in which the detector response has fluctuated 
upwards by several standard deviations and/or the spectator system 
accompanying the hard scattering process, including the initial state 
radiation, makes an unusually large contribution to the \sumet. 
We have set the minimum $p_T^{hard}$ to 60 GeV/$c$. The contribution to 
the \sumet $> 420$~GeV Monte Carlo sample from events with 
$p_T^{hard} < 60$~GeV/$c$ is estimated to be less than $1\%$. 

\end{subsection}
\begin{subsection} {The NJETS QCD Matrix Element Calculation}

The NJETS Monte Carlo program \cite{NJETS} 
provides parton-level predictions based on the 
LO QCD $2 \rightarrow N$ matrix elements. 
The calculation requires the minimum separation 
between the final state partons in ($\eta,\phi$)-space 
to exceed $\Delta R_{MIN}$. We have set $\Delta R_{MIN} = 0.9$, 
and have used the KMRSD0 structure 
function parameterization~\cite{kmrsd} with the renormalization scale 
chosen to be the average $p_T$ of the outgoing partons. 
NJETS does not use a parton fragmentation model. 
Jet definitions and selection cuts are therefore applied to the 
final state partons. 
To enable a direct comparison between NJETS predictions and observed 
distributions we have 
smeared the final state parton energies in our NJETS calculations 
with the Gaussian jet energy resolution function:
\begin{equation}
 \sigma_{E} = 0.1 \; E \; . 
\end{equation}
This provides a good approximation to the CDF jet resolution function. 

\end{subsection}
\begin{subsection} {Phase-Space Model}

We have used the GENBOD phase-space generator~\cite{cernlib} to generate 
samples of Monte Carlo events for which the 
multijet systems uniformly populate the N-body phase-space. 
These phase-space Monte Carlo 
events were generated with single-jet masses distributed according to the 
single-jet mass distribution predicted 
by the HERWIG Monte Carlo program. In addition, the multijet mass distributions 
were generated according to the corresponding distributions obtained 
from the HERWIG Monte Carlo calculation. 
Comparisons between the resulting phase-space model distributions and 
the corresponding HERWIG and NJETS Monte Carlo distributions help us to 
understand which multijet parameters are most sensitive to the 
behaviour of QCD multijet matrix elements. 

\end{subsection}
\end{section}

\begin{section} {Multijet Variables}
To completely specify a system of N jets in the N-jet rest-frame 
we require (4N-3) independent parameters. However, the N-jet system 
can be rotated about the beam direction without losing any 
interesting information. Hence we need only specify (4N-4) 
parameters. We will use the N-jet mass and 
the (4N-5) dimensionless variables introduced and discussed 
in Ref.~\cite{preprint}. In the following the variables 
are briefly reviewed.

\begin{subsection} {Three-Jet Variables}

In previous three-jet analyses 
\cite{ua1_3jet,cdf_3jet,d0_34jet} 
it has become traditional to label 
the incoming interacting partons 1 and 2, and 
the outgoing jets 3, 4, and 5, with the jets ordered such that 
$E_3 > E_4 > E_5$, where $E_j$ is the energy of jet $j$ in the three-body 
rest-frame. At fixed three-jet mass $m_{3J}$ a system of three massless 
jets can be specified in the three-jet rest-frame using 
four dimensionless variables, $X_3, X_4, \cos \theta_3$, and $\psi_3$, 
which are defined:
\begin{itemize}
\item[(i)] The Dalitz variables $X_3$ and $X_4$:
\begin{eqnarray}
X_j & \equiv & \frac{2\;E_j}{m_{3J}} \; .
\end{eqnarray}
\item[(ii)] The cosine of the leading jet scattering angle:
\begin{equation}
  \cos\theta_3  \; \equiv \; \frac{\overrightarrow{P}\!\!_{AV}\cdot
                      \overrightarrow{P}\!_{3}}
                   {| \overrightarrow{P}\!\!_{AV}|
             | \overrightarrow{P}\!_{3} |} \; , \\
\end{equation}
where the average beam direction:
\begin{eqnarray}
\overrightarrow{P}\!\!_{AV} & = & \overrightarrow{P}\!_1 
                                - \overrightarrow{P}\!_2 \; ,
\end{eqnarray}
and particle 1 is the incoming interacting parton with the highest 
energy in the laboratory frame.\\
\item[(iii)] $\psi_3$, defined in the three-jet rest-frame as
the angle between the three-jet plane and the plane containing the 
leading-jet and the average beam direction:
\begin{eqnarray}
  \cos\psi_{3}  & \equiv & 
\frac{{(\overrightarrow{P}\!_{3} \times 
      \overrightarrow{P}\!\!_{AV})} \cdot
      {(\overrightarrow{P}\!_{4} \times \overrightarrow{P}\!_{5})}}
{| {\overrightarrow{P}\!_{3} \times \overrightarrow{P}\!\!_{AV}}|
| {\overrightarrow{P}\!_{4} \times \overrightarrow{P}\!_{5}}|} \; .
\end{eqnarray}
\end{itemize}
To specify a system of three massive jets we must supplement 
the traditional three-jet variables with three additional parameters that 
describe the single-jet masses. These parameters are taken to be the 
single-jet mass fractions $f_3, f_4$, and $f_5$, where:
\begin{eqnarray}
f_j & \equiv & \frac{m_j}{m_{3J}} \; .
\end{eqnarray}
Thus we have eight three-jet variables which consist of $m_{3J}$,  
four parameters that specify the three-jet configuration
($X_3, X_4, \cos \theta_3$, and $\psi_3$), 
and three variables that specify the single-jet masses 
($f_3, f_4$, and $f_5$). 
\end{subsection}

\begin{subsection} {Four-Jet and Five-Jet Variables}

A multijet system with more than three-jets can be partially specified 
using the three-jet variables described above. This is accomplished by 
first reducing the multijet system to a three-body system. A four-jet 
system is reduced to a three-body system by combining the two-jets with 
the lowest two-jet mass. The resulting three-body system can then be described 
using the variables $m_{4J}$, 
$X_{3'}, X_{4'}, \cos \theta_{3'}$, $\psi_{3'}$, 
$f_{3'}, f_{4'}$, and $f_{5'}$, 
where the primes remind us that two jets have been combined. 
A five-jet system is reduced to a three-body system by first combining the 
two-jets with the lowest two-jet mass to obtain a four-body system, and 
then combining the two bodies with the lowest two-body mass to obtain a 
three-body system. The resulting three-body system can then be described 
using the variables $m_{5J}$, 
$X_{3''}, X_{4''}, \cos \theta_{3''}$, $\psi_{3''}$, 
$f_{3''}, f_{4''}$, and $f_{5''}$.

To complete the description of four-jet (five-jet) events we must now 
specify a further four (eight) variables that describe how the multijet 
system has been reduced to a three-body system. Consider first the step 
in which a four-body system is reduced to a three-body system. We label 
the two objects being combined A and B with $E_A > E_B$, where $E_A$ and $E_B$ 
are energies in the four-body rest-frame. 
To describe the (AB)-system we use the following four variables:
\begin{itemize}
\item[(a)] The normalized masses $f_{A}$ and $f_{B}$:
\begin{eqnarray}
f_{A} \; \equiv \; \frac{m_{A}}{m_{NJ}} & {\rm and} &  
f_{B} \; \equiv \;  \frac{m_{B}}{m_{NJ}} \; ,
\end{eqnarray}
where $m_{NJ}$ is the mass of the multijet system.
\item[(b)] The two-body energy sharing variable $X_{A}$, 
defined in the multijet rest-frame as the fraction 
of the energy of the (AB)-system taken by A:
\begin{eqnarray}
X_{A}  & \equiv & \frac{E_A}{E_A + E_B} \; .
\end{eqnarray}
\item[(c)] The two-body angular variable $\psi'_{AB}$, 
defined in the multijet rest-frame as the 
angle between (i) the plane containing the ($AB$)-system and the average
beam direction, and (ii) the plane containing A and B. 
The prime reminds us that in order to define $\psi'_{AB}$ 
we have combined two bodies to obtain the (AB)-system. Note that:
\begin{eqnarray}
  \cos\psi'_{AB}  & \equiv & \frac{(\overrightarrow{P}\!_{A} \times
\overrightarrow{P}\!_{B}) \cdot
(\overrightarrow{P}\!\!_{AB} \times \overrightarrow{P}\!\!_{AV})}
{| \overrightarrow{P}\!_{A} \times \overrightarrow{P}\!_{B} |
 | \overrightarrow{P}\!\!_{AB} \times \overrightarrow{P}\!\!_{AV} |} \; .
\end{eqnarray}
\end{itemize}
For five-jet events we must also specify the step in which the five-jet 
system is reduced to a four-body system.  We label the two jets that are 
combined C and D with $E_C > E_D$, and use the four variables 
$f_C$, $f_D$, $X_C$, and $\psi''_{CD}$.

In summary, a four-jet system is described using 12 variables: $m_{4J}$,  
$X_{3'}$, $X_{4'}$, $\cos \theta_{3'}$, $\psi_{3'}$, 
$f_{3'}$, $f_{4'}$, $f_{5'}$, 
$f_A$, $f_B$, $X_A$, and $\psi'_{AB}$.
A five-jet system is described using 16 variables: $m_{5J}$, 
$X_{3''}$, $X_{4''}$, $\cos \theta_{3''}$, $\psi_{3''}$, 
$f_{3''}$, $f_{4''}$, $f_{5''}$,
$f_{A'}$, $f_{B'}$, $X_{A'}$, $\psi''_{A'B'}$,
$f_C$, $f_D$, $X_C$, and $\psi''_{CD}$. 
Note that following the convention of Ref.~\cite{preprint} the 
primes indicate which parameters are defined after one or two 
steps in which two objects have been combined. 

\end{subsection} 
\end{section}

\begin{figure}
\parbox[t]{6.0in}{
\hspace{-1cm}
\vspace{-0.5in}
\begin{center}
\leavevmode
\epsfysize=6.0in
\epsffile[20 143 575.75 698.75]{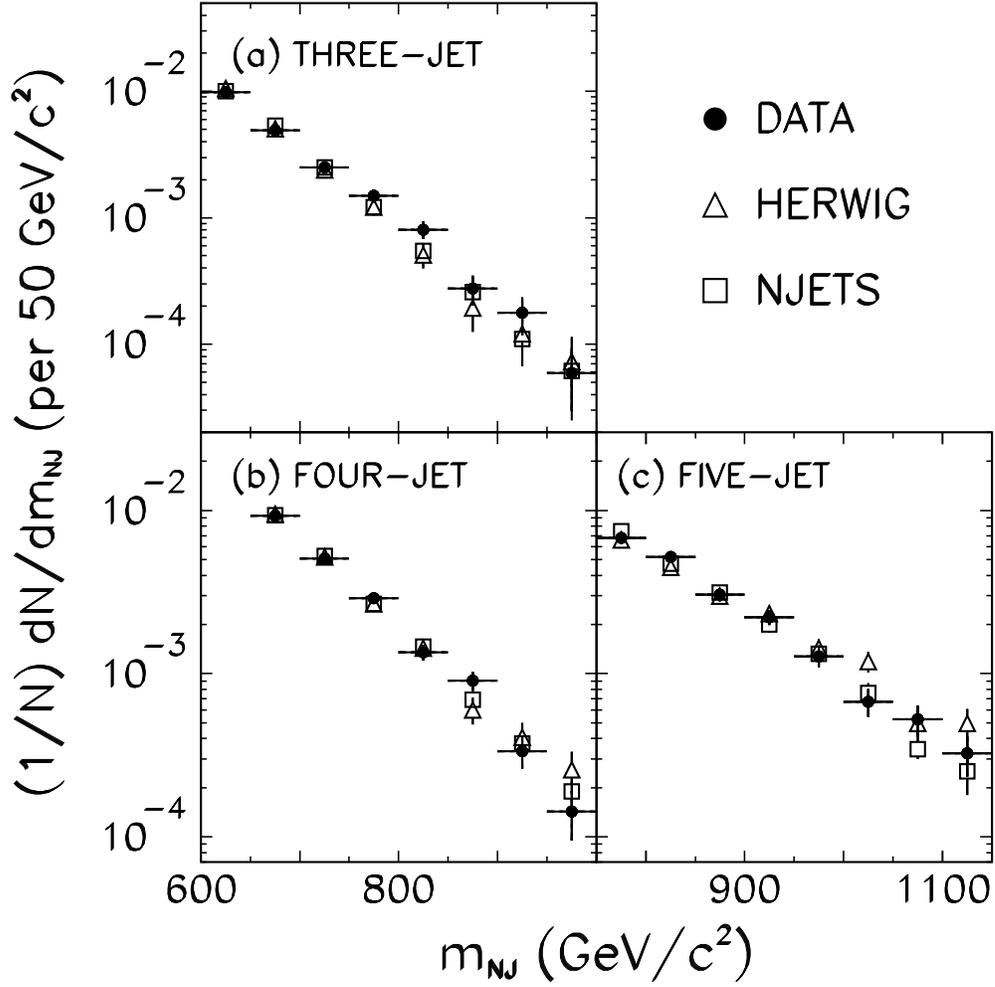}
\end{center}
}
\caption{Inclusive multijet mass distributions for topologies with 
(a) three jets, (b) four jets, and (c) five jets. 
Observed distributions (points) 
are compared with HERWIG predictions (triangles) and NJETS predictions
(squares).}
\label{fig:mnj}
\end{figure}

\begin{figure}
\parbox[t]{6.0in}{
\hspace{-1cm}
\vspace{-0.5in}
\begin{center}
\leavevmode
\epsfysize=6.0in
\epsffile[20 143 575.75 698.75]{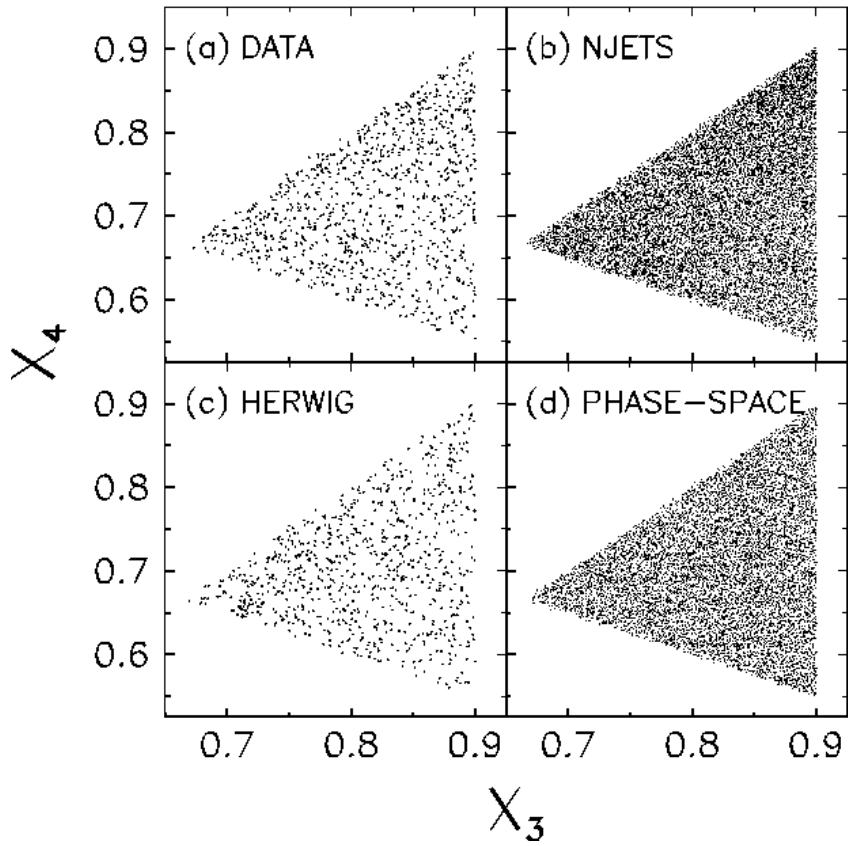}
\end{center}
}
\caption{Three-jet Dalitz distributions after imposing the requirements 
$m_{3J} > 600$~GeV/$c^2$, $X_{3} < 0.9$, and $| \cos\theta_{3}| < 0.6$, 
shown for (a) data, (b) NJETS, (c) HERWIG, and (d) the phase-space model.} 
\label{fig:3jet_x3_vs_x4}
\end{figure}

\begin{figure}
\parbox[t]{6.0in}{
\hspace{-1cm}
\vspace{-1.5in}
\begin{center}
\leavevmode
\epsfysize=6.0in
\epsffile[20 143 575.75 698.75]{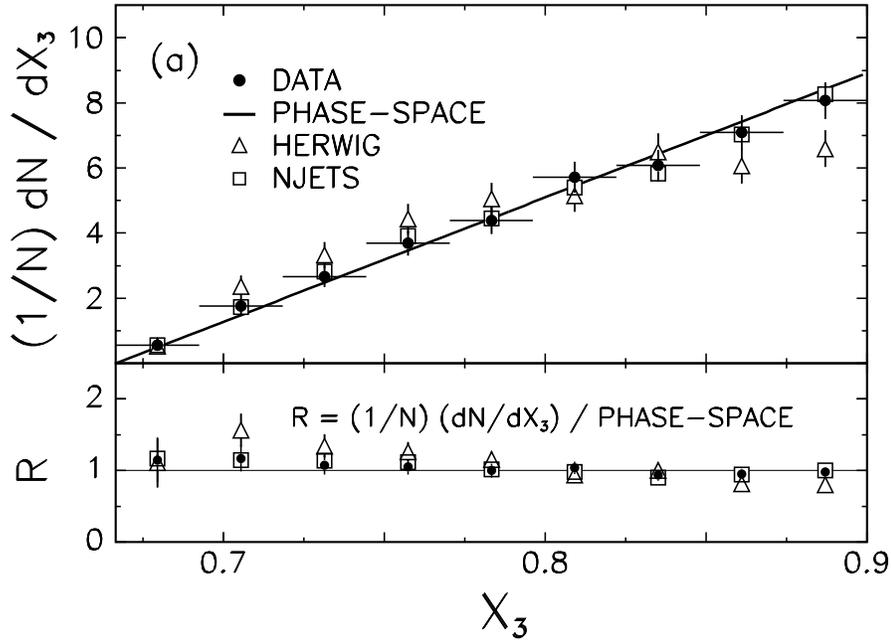}
\end{center}
}
\parbox[b]{6.0in}{
\hspace{-1cm}
\vspace{-2.6in}
\begin{center}
\leavevmode
\epsfysize=6.0in
\epsffile[20 143 575.75 698.75]{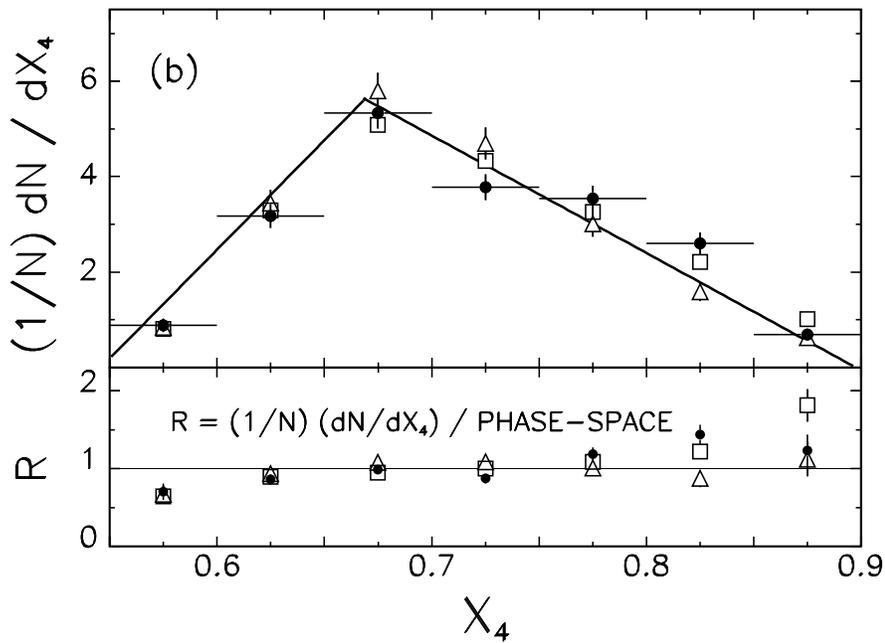}
\end{center}
}
\vspace{-1.5in}
\caption{Inclusive three-jet Dalitz distributions for events that 
satisfy the requirements 
$m_{3J} > 600$ GeV/$c^2$, $X_{3} < 0.9$, and $| \cos\theta_{3}| < 0.6$.
Data (points) are compared with HERWIG predictions (triangles), NJETS 
predictions (squares), and phase-space model predictions (curves) for 
(a) $X_{3}$, and (b) $X_{4}$.}
\label{fig:3jet_x3_x4}
\end{figure}

\begin{figure}
\parbox[t]{6.0in}{
\hspace{-1cm}
\vspace{-0.5in}
\begin{center}
\leavevmode
\epsfysize=6.0in
\epsffile[20 143 575.75 698.75]{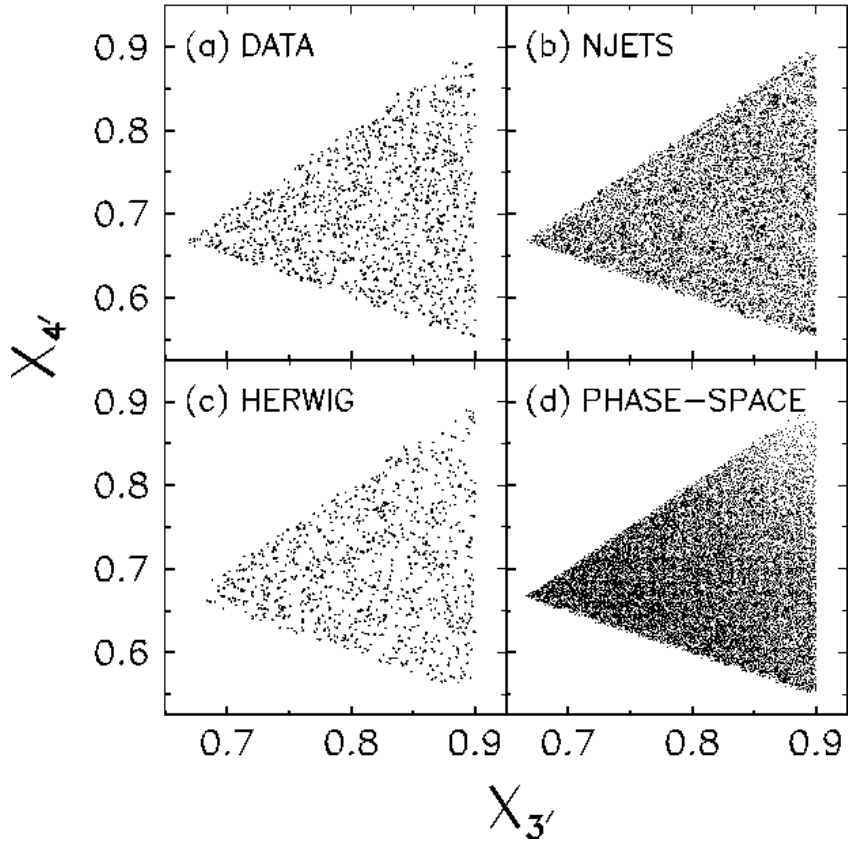}
\end{center}
}
\caption{Inclusive four-jet Dalitz distributions for events that 
that satisfy the requirements 
$m_{4J} > 650$ GeV/$c^2$, $X_{3'} < 0.9$, and $| \cos\theta_{3'}| < 0.8$, 
shown 
for (a) data, (b) NJETS, (c) HERWIG, and (d) phase-space model predictions.}
\label{fig:4jet_x3_vs_x4}
\end{figure}

\begin{figure}
\parbox[t]{6.0in}{
\hspace{-1cm}
\vspace{-1.5in}
\begin{center}
\leavevmode
\epsfysize=6.0in
\epsffile[20 143 575.75 698.75]{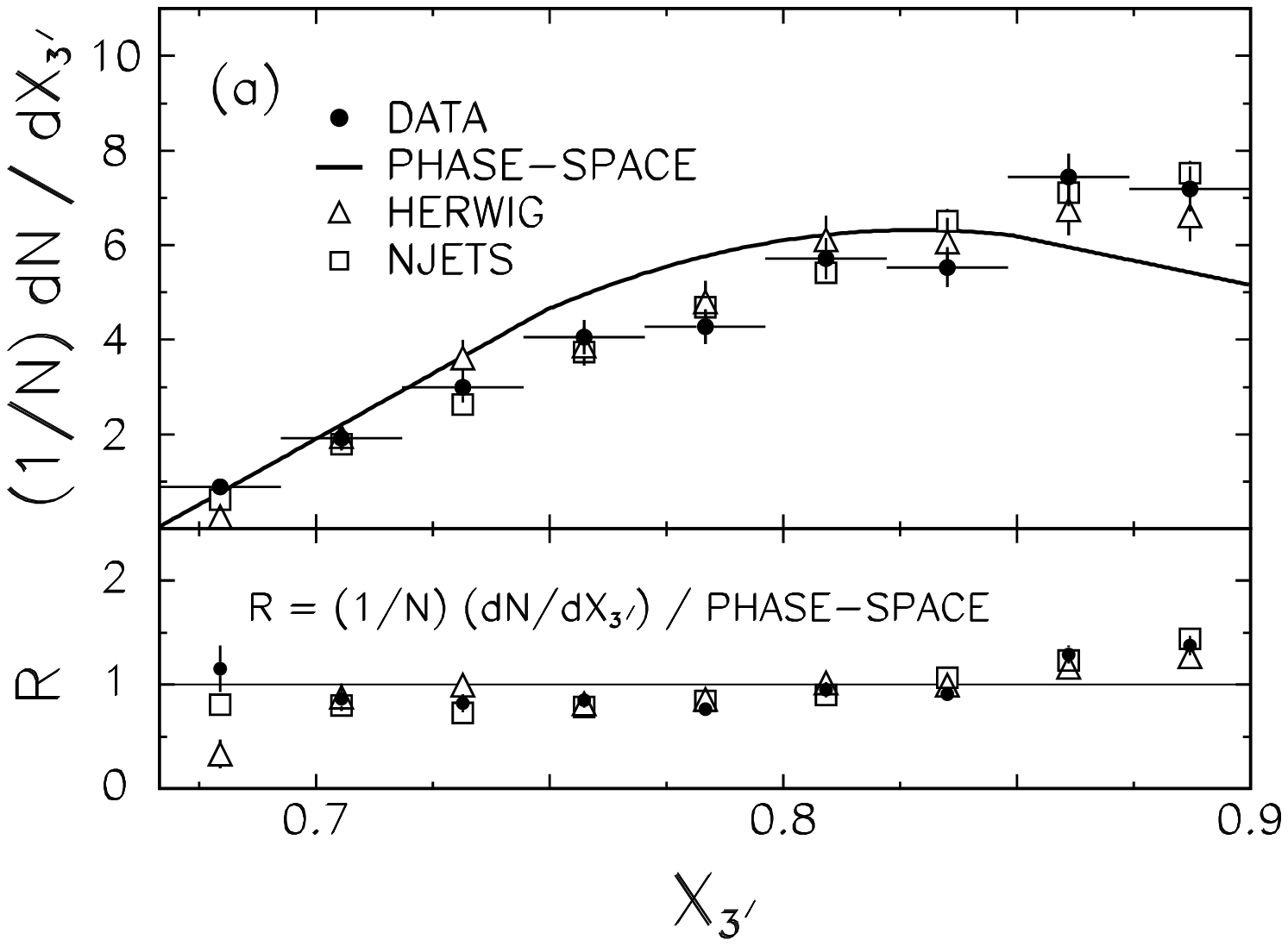}
\end{center}
}
\parbox[b]{6.0in}{
\hspace{-1cm}
\vspace{-2.6in}
\begin{center}
\leavevmode
\epsfysize=6.0in
\epsffile[20 143 575.75 698.75]{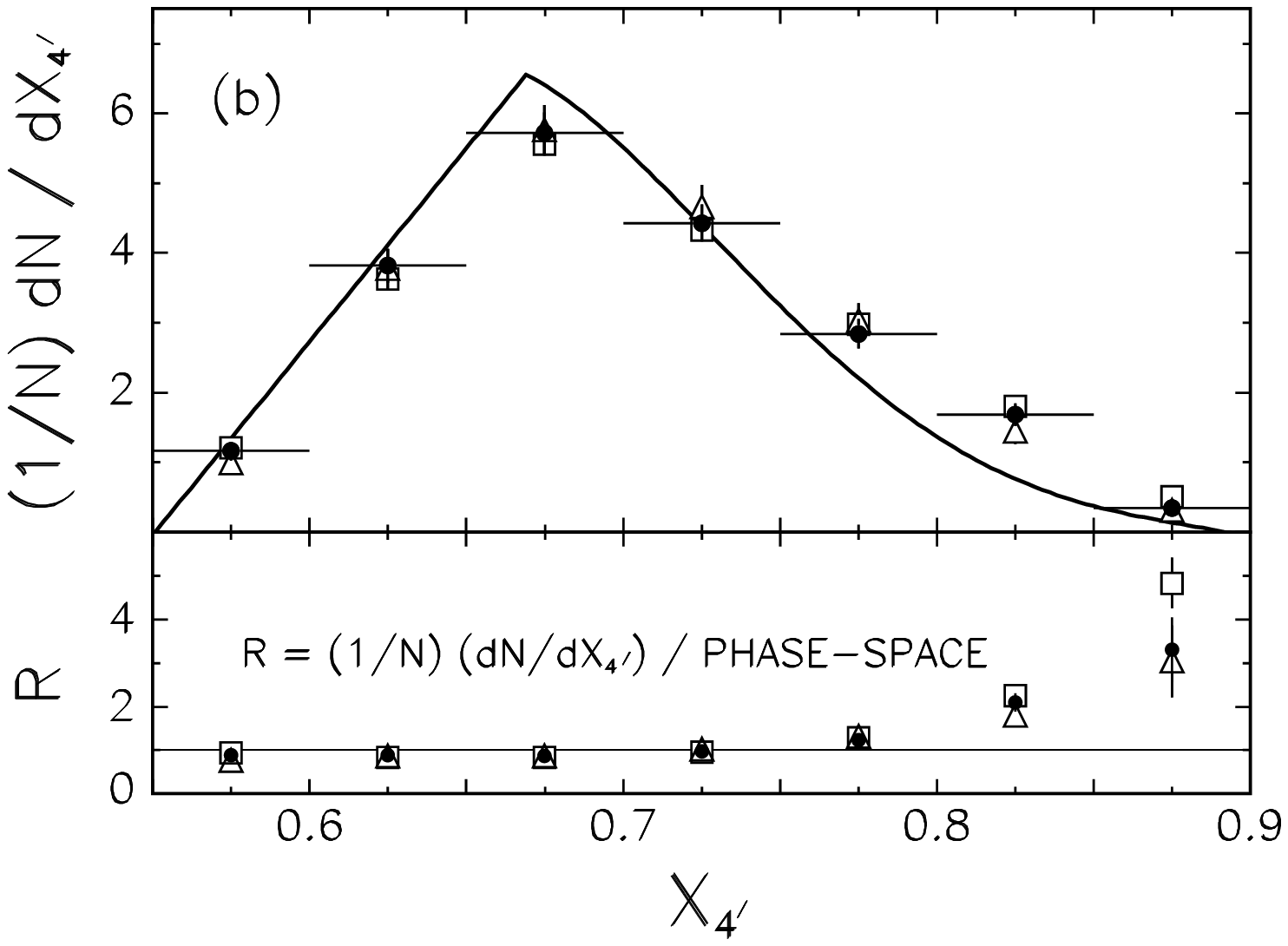}
\end{center}
}
\vspace{-1.5in}
\caption{Dalitz distributions for inclusive four-jet topologies 
that satisfy the requirements 
$m_{4J} > 650$ GeV/$c^2$, $X_{3'} < 0.9$, and $| \cos\theta_{3'}| < 0.8$.
Data (points) are compared with HERWIG predictions (triangles), NJETS 
predictions (squares), and phase-space model predictions (curves) for 
(a) $X_{3'}$, and 
(b) $X_{4'}$.}
\label{fig:4jet_x3_x4}
\end{figure}

\begin{figure}
\parbox[t]{6.0in}{
\hspace{-1cm}
\vspace{-0.5in}
\begin{center}
\leavevmode
\epsfysize=6.0in
\epsffile[20 143 575.75 698.75]{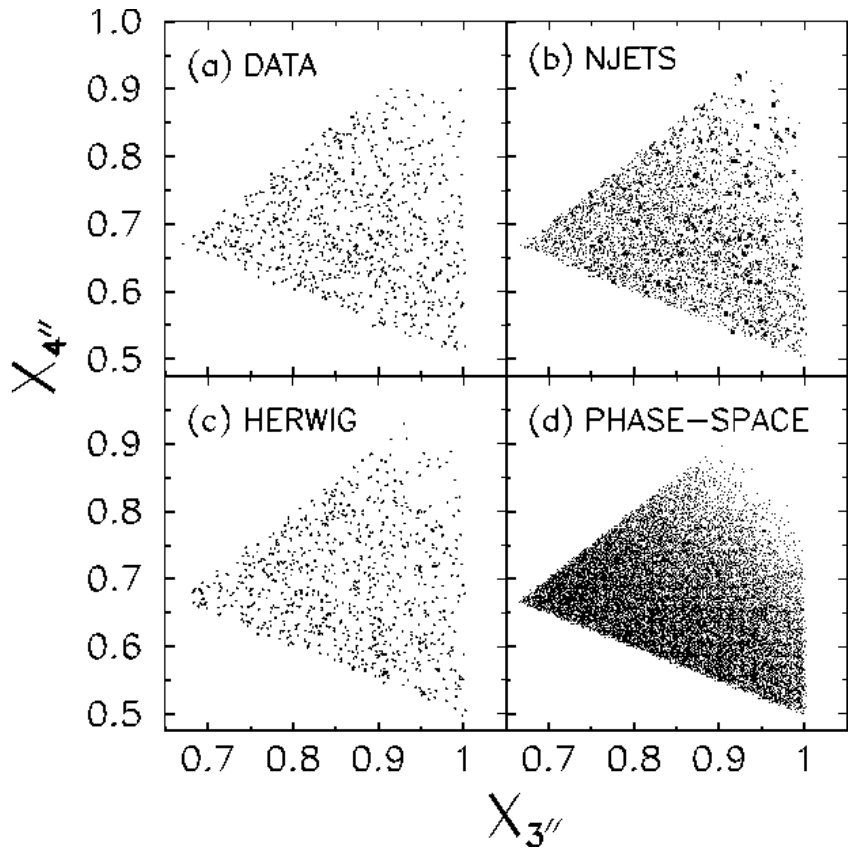}
\end{center}
}
\caption{Inclusive five-jet Dalitz distributions for events 
that satisfy the requirement 
$m_{5J} > 750$ GeV/$c^2$, shown
for (a) data, (b) NJETS, (c) HERWIG, and (d) phase-space model predictions.}
\label{fig:5jet_x3_vs_x4}
\end{figure}

\begin{section} {Results}

The (4N-4) multijet variables described in the previous sections provide 
a set of independent parameters that span the multijet parameter space 
in the multijet rest frame. In the following the (4N-4) 
multijet distributions are compared with QCD and phase-space model predictions. 
All distributions are inclusive. 
If there are more than N jets in an event, the 
N highest $E_T$ jets are used to define the multijet system. 
It should be noted that at fixed multijet mass the $\sum E_T > 420$~GeV, 
$\Delta R \geq 0.9$, and $E_T > 20$~GeV 
requirements place restrictions on the available multijet parameter 
space. Consequently, some regions of parameter space are depopulated due 
to a low experimental acceptance. These inefficient regions can be largely 
avoided in the three-jet analysis by placing suitable requirements on the 
multijet mass, leading-jet angle, and leading-jet Dalitz variable. In the 
following we have required that 
$m_{3J} > 600$~GeV/$c^2$, $|\cos \theta_3| < 0.6$, and $X_3 < 0.9$. 
These requirements select 1021 events with three-or-more jets, 
of which 320 events have more than three jets. 
Events entering the inclusive four-jet distributions are required to have 
$m_{4J} > 650$~GeV/$c^2$, $|\cos \theta_{3'}| < 0.8$, and $X_{3'} < 0.9$. 
These requirements select 1273 events with four-or-more jets, of which 
245 events have more than four jets. Only 226 events enter into both 
the inclusive three-jet and inclusive four-jet distributions..  
Note that the four-jet requirements reduce, but do not completely eliminate, 
the regions of low experimental acceptance. A more restrictive $X_{3'}$ 
requirement could be used to remove events populating the remaining region 
of low acceptance, but would cost a large reduction in statistics. 
Given the limited statistics of the present data sample, we have chosen to 
tolerate some regions of low experimental acceptance and use the phase-space 
model predictions to understand which regions of parameter space are 
affected. 
Finally, the inclusive five-jet data sample has very limited statistics, 
and we have therefore chosen to apply only the requirement 
$m_{5J} > 750$~GeV/$c^2$ to events entering the five-jet distributions.
This requirement selects 817 events with five-or-more jets, of which 
146 events have more than five jets. Only 148 events enter into both 
the five-jet and four-jet distributions, and only 42 events enter into 
both the five-jet and three-jet distributions.

\begin{subsection} {Multijet Mass Distributions}

In Ref.~\cite{multijet_paper} HERWIG and NJETS QCD calculations were 
shown to give a good description of the shapes of the observed 
multijet mass distributions for exclusive samples of high-mass multijet 
events. In Figs.~\ref{fig:mnj}a, \ref{fig:mnj}b, and \ref{fig:mnj}c both the 
HERWIG and NJETS predictions are shown to give good descriptions of the 
shapes of the inclusive $m_{3J}$, $m_{4J}$, and $m_{5J}$ distributions for the 
high-mass multijet samples described in this paper. Note that over the 
limited mass range of the present data sample, to a good approximation 
the $m_{NJ}$ distributions 
are falling exponentially with increasing mass. 

\end{subsection}

\begin{subsection} {Three-Body Dalitz Distributions}

We begin by considering the inclusive three-jet Dalitz distributions. 
Event populations in the ($X_3$, $X_4$)-plane are shown in 
Fig.~\ref{fig:3jet_x3_vs_x4} 
for (a) data, (b) NJETS, (c) HERWIG, and (d) phase-space model predictions. 
The phase-space population is uniform over the kinematically allowed region. 
Neither the data nor the QCD predictions exhibit large density variations in 
the ($X_3$, $X_4$)-plane in the region of interest ($X_3 < 0.9$), although 
with the relatively high statistical precision of the NJETS predictions 
the tendency for the predicted event density to increase as $X_4$ 
becomes large is visible (note that as $X_4 \rightarrow 1$ 
the third-to-leading jet Dalitz variable $X_5 \rightarrow 0$). 
The observed $X_3$ distribution is compared with phase-space model and QCD 
predictions in Fig.~\ref{fig:3jet_x3_x4}a. The corresponding 
comparisons for the $X_4$ distribution are shown in 
Fig.~\ref{fig:3jet_x3_x4}b. The HERWIG and NJETS predictions 
give reasonable descriptions of the observed distributions. 
Note that the observed distributions are not very different from 
the phase-space model predictions. 

We next consider the inclusive four-jet distributions. 
Event populations in the ($X_{3'}$, $X_{4'}$)-plane are shown in 
Fig.~\ref{fig:4jet_x3_vs_x4} 
for (a) data, (b) NJETS, (c) HERWIG, and (d) phase-space model predictions. 
The phase-space population is not uniform over the kinematically allowed 
region. Care must therefore be taken in interpreting the 
distributions. The data and the QCD predictions exhibit a more 
uniform event density over the ($X_{3'}$, $X_{4'}$)-plane. 
The observed $X_{3'}$ distribution is compared with phase-space model and QCD 
predictions in Fig.~\ref{fig:4jet_x3_x4}a. The corresponding 
comparisons for the $X_{4'}$ distribution are shown in 
Fig.~\ref{fig:4jet_x3_x4}b. The HERWIG and NJETS predictions 
give reasonable descriptions of the observed distributions. 
Note that compared to the phase-space model predictions, the data and QCD 
predictions prefer topologies with large $X_{3'}$ and large $X_{4'}$ 
(note that as $X_{3'} \rightarrow 1$ 
the three-body topology approaches a two-body configuration, and 
as $X_{4'} \rightarrow 1$ we have $X_{5'} \rightarrow 0$). 

Finally, consider the inclusive five-jet distributions. 
Event populations in the ($X_{3''}$, $X_{4''}$)-plane are shown in 
Fig.~\ref{fig:5jet_x3_vs_x4} 
for (a) data, (b) NJETS, (c) HERWIG, and (d) phase-space model predictions. 
Again, the phase-space population is not uniform over the kinematically allowed 
region, and care must be taken in interpreting the 
distributions. The observed event population and the QCD predictions are more 
uniformly distributed over the ($X_{3''}$, $X_{4''}$)-plane. 
However, all distributions 
are depleted as  $X_{3''} \rightarrow 1$ and $X_{4''} \rightarrow 1$. 
The observed $X_{3''}$ distribution is compared with phase-space model and QCD 
predictions in Fig.~\ref{fig:5jet_x3_x4}a. The corresponding 
comparisons for the $X_{4''}$ distribution are shown in 
Fig.~\ref{fig:5jet_x3_x4}b. The HERWIG and NJETS predictions 
give reasonable descriptions of the observed distributions. 
Note that compared to the phase-space model predictions, the data and QCD 
predictions prefer topologies with $X_{3''} \rightarrow 1$ and 
$X_{4''} \rightarrow 1$. 

\end{subsection}

\begin{subsection} {Three-Body Angular Distributions}

We begin by considering the inclusive three-jet angular distributions. 
Event populations in the ($\cos \theta_3$, $\psi_3$)-plane are shown in 
Fig.~\ref{fig:3jet_cos_vs_psi} 
for (a) data, (b) NJETS, (c) HERWIG, and (d) phase-space model predictions. 
The phase-space population is approximately uniform. 
In contrast both the observed distribution and the QCD predictions exhibit 
large density variations over the ($\cos \theta_3$, $\psi_3$)-plane, with 
the event density increasing as $|\cos\theta_3| \rightarrow 1$ and 
$\psi_3 \rightarrow 0$ or $\pi$. 
The increase in event rate as $|\cos\theta_3| \; \rightarrow 1$ is 
similar to the behaviour of the leading-jet angular distribution 
resulting from the $2 \rightarrow 2$ LO QCD matrix element. The increase in 
event rate as $\psi_3 \rightarrow 0$ or $\pi$ reflects 
the preference of the three-jet matrix element for topologies which are 
planar. It is interesting 
to note that as $\cos\theta_3 \rightarrow 1$ the NJETS calculation 
shows a preference for configurations with $\psi_3 \rightarrow 0$ 
rather than $\pi$ and as  $\cos\theta_3 \rightarrow -1$ the NJETS calculation 
shows a preference for configurations with $\psi_3 \rightarrow \pi$ 
rather than 0. These preferred regions of the 
parameter space correspond to configurations in which jet 5 is closer to 
the beam direction, and therefore reflect the initial state radiation pole 
in the matrix element.
The observed $\cos\theta_3$ distribution is compared with phase-space model 
and QCD predictions in Fig.~\ref{fig:3jet_cos_psi}a. 
The corresponding 
comparisons for the $\psi_3$ distribution are shown in 
Fig.~\ref{fig:3jet_cos_psi}b. 
Both HERWIG 
and NJETS predictions 
give reasonable descriptions of the observed distributions, which are very  
different from the phase-space model predictions. Note that the observed 
$\cos\theta_3$ distribution is also very similar to the LO prediction 
for $q\overline{q} \rightarrow q\overline{q}$ scattering~\cite{combridge}.

Next, consider the inclusive four-jet angular distributions. 
Event populations in the ($\cos \theta_{3'}$, $\psi_{3'}$)-plane are shown in 
Fig.~\ref{fig:4jet_cos_vs_psi} 
for (a) data, (b) NJETS, (c) HERWIG, and (d) phase-space model predictions. 
The phase-space population is approximately uniform. 
In contrast both the observed distribution and the QCD predictions exhibit 
large density variations over the ($\cos \theta_{3'}$, $\psi_{3'}$)-plane, with 
the event density increasing as $|\cos\theta_{3'}| \rightarrow 1$ and 
$\psi_{3'} \rightarrow 0$ or $\pi$. 
This behaviour is similar to the behaviour of the corresponding three-jet 
distributions. 
The observed $\cos\theta_{3'}$ distribution is compared with phase-space model 
and QCD predictions in Fig.~\ref{fig:4jet_cos_psi}a. 
The corresponding 
comparisons for the $\psi_{3'}$ distribution are shown in 
Fig.~\ref{fig:4jet_cos_psi}b. Both HERWIG and NJETS predictions 
give reasonable descriptions of the observed distributions, which are very  
different from the phase-space model predictions. Note that the observed 
$\cos\theta_{3'}$ distribution is also very similar to the LO prediction 
for $q\overline{q} \rightarrow q\overline{q}$ scattering.

Finally, consider the inclusive five-jet angular distributions. 
Event populations in the ($\cos \theta_{3''}$, $\psi_{3''}$)-plane are shown in 
Fig.~\ref{fig:5jet_cos_vs_psi} 
for (a) data, (b) NJETS, (c) HERWIG, and (d) phase-space model predictions. 
The phase-space population is not uniform, and care must therefore be taken 
in interpreting the distributions. 
However, both the observed distribution and the QCD predictions exhibit 
much larger density variations over the 
($\cos \theta_{3''}$, $\psi_{3''}$)-plane, with 
the event density increasing as $|\cos\theta_{3''}| \rightarrow 1$ and 
$\psi_{3''} \rightarrow 0$ or $\pi$. 
This behaviour is similar to the behaviour of the corresponding three-jet 
distributions. 
The observed $\cos\theta_{3''}$ distribution is compared with phase-space model 
and QCD predictions in Fig.~\ref{fig:5jet_cos_psi}a. The corresponding 
comparisons for the $\psi_{3''}$ distribution are shown in 
Fig.~\ref{fig:5jet_cos_psi}b. Both HERWIG and NJETS predictions 
give reasonable descriptions of the observed distributions, which are very  
different from the phase-space model predictions. Note that the observed 
$\cos\theta_{3''}$ distribution is also very similar to the LO prediction 
for $q\overline{q} \rightarrow q\overline{q}$ scattering even though there 
are now five jets in the final state.

\end{subsection}

\begin{subsection} {Single-Body Mass Distributions for Three-Body Systems}

The single-jet mass fraction distributions are shown in 
Fig.~\ref{fig:3jet_f345} for inclusive three-jet events. The $f_j$ 
distributions are reasonably well described by the HERWIG Monte Carlo 
predictions, although there is a tendency for the HERWIG fragmentation 
model to slightly overestimate the fraction of low-mass jets. 
The observed distributions 
peak at $f_j \sim 0.05$ or less. Hence, for many purposes, 
jets at high energy can be considered to be massless.  Note that 
since jets are massless in the matrix element calculations, there 
are no NJETS predictions for the $f_j$ distributions. 

The $f_{j'}$ and $f_{j''}$ distributions are shown for 
inclusive four-jet and inclusive five-jet 
events in Figs.~\ref{fig:4jet_f345} and \ref{fig:5jet_f345} respectively. 
These distributions 
exhibit a single-jet peak at low mass-fractions (less than 0.05), 
and have a long tail associated with two-jet $j'$ systems, and 
two-jet or three-jet $j''$ systems. The HERWIG predictions give a good 
description of all the distributions except perhaps at very low mass fractions 
(less than 0.05) where there is tendency to overestimate the observed 
jet rate. Although the NJETS calculations do not provide predictions for 
the single-jet part of the $f_{j'}$ and $f_{j''}$ distributions, they are seen 
to correctly predict the tail associated with multijet $j'$ and $j''$ systems. 

\end{subsection}

\begin{subsection} {Two-Body Energy Sharing Distributions}

The observed $X_A$ distribution is shown in 
Figs.~\ref{fig:45jet_xa_xc}a and \ref{fig:45jet_xa_xc}b 
to be reasonably well described by the HERWIG and NJETS predictions. 
The data and the QCD predictions 
favor a more asymmetric sharing of energy between the two jets A and B than 
predicted by the phase-space model. This reflects the presence 
of the soft gluon radiation pole in the QCD matrix element. In 
Figs.~\ref{fig:45jet_xa_xc}c and \ref{fig:45jet_xa_xc}d the $X_{A'}$ 
distributions are shown to be 
qualitatively similar to the corresponding $X_A$ distributions, and also 
similar to the corresponding $X_C$ distributions shown in 
Figs.~\ref{fig:45jet_xa_xc}e and \ref{fig:45jet_xa_xc}f. 
In general the data are reasonably well described by
 the QCD predictions and are very different from the phase-space model 
predictions.

\end{subsection}

\begin{subsection} {Two-Body Angular Distributions}

The observed $\psi'_{AB}$ distribution is shown in 
Figs.~\ref{fig:45jet_pab_pcd}a and \ref{fig:45jet_pab_pcd}b to be 
well described by the HERWIG and NJETS predictions. 
The phase-space model prediction is also approximately uniform, but 
underestimates the fraction of events in which 
the plane of the two-body system is close to the plane containing the 
two-body system and the beam direction ($\psi'_{AB} \rightarrow 0$ or $\pi$). 
In Figs.~\ref{fig:45jet_pab_pcd}c and \ref{fig:45jet_pab_pcd}d the 
$\psi''_{A'B'}$ distributions are shown to be 
qualitively similar to the corresponding $\psi'_{AB}$ distributions. 
The $\psi''_{CD}$ distributions shown in Figs.~\ref{fig:45jet_pab_pcd}e and 
\ref{fig:45jet_pab_pcd}f are similar to the phase-space model 
predictions. In all cases the data are well described by the QCD 
predictions. None of the observed distributions are very different from 
the phase-space model predictions, 
although the phase-space model calculation does underestimate the event rate 
as $\psi'_{AB} \rightarrow 0$ or $\pi$, or 
as $\psi''_{A'B'} \rightarrow 0$ or $\pi$. 

\end{subsection}
\begin{subsection} {Single-Body Mass Distributions for Two-Body Systems}

The observed $f_A$, $f_B$, $f_{A'}$, $f_{B'}$, $f_C$, and $f_D$ distributions 
are shown in Figs.~\ref{fig:45jet_fabcd}a, \ref{fig:45jet_fabcd}b, 
\ref{fig:45jet_fabcd}c, \ref{fig:45jet_fabcd}d, \ref{fig:45jet_fabcd}e and 
\ref{fig:45jet_fabcd}f respectively to be reasonably well described 
by the HERWIG predictions although there is a tendency for the HERWIG 
predictions to overestimate the jet rate at very small single-jet masses. 
In all cases the distributions exhibit a single-jet 
mass peak at small mass fractions ($\sim 0.02$ or less). The  $f_{A'}$ and 
$f_{B'}$ have a long high-mass tail which corresponds to two-jet $A'$ and $B'$ 
systems. This tail is well described by the NJETS predictions.

\end{subsection}

\begin{subsection}{$\chi^2$ Test}

In general both NJETS and HERWIG predictions give a 
good first description of the observed multijet distributions, which 
correspond to (4N-4) variables that span the N-body parameter space. 
A more quantitative assessment can be made by examining the $\chi^2$ 
per degree of freedom that characterizes the agreement between the observed 
distributions and the QCD predictions. The $\chi^2$ are listed for each 
distribution in Table 1. The computed $\chi^2$'s take into account 
statistical uncertainties on both measured points and the QCD Monte Carlo 
predictions, but do not take into account systematic uncertainties. 
In Ref.~\cite{multijet_paper} 
the systematic uncertainties were mapped out for a limited set of multijet 
distributions, and found to be small compared to statistical uncertainties. 
Unfortunately, for the more complicated multijet parameter space of the 
present analysis, limited computing resources do not allow us to fully 
map out the systematic uncertainties on the predictions. 
However, even in the absence of a full evaluation of the systematic 
uncertainties, an examination of Table 1 shows that 
NJETS provides a reasonable description of all of the observed multijet 
distributions except perhaps the $X_A$ distribution. 
The combined $\chi^2$ for the NJETS description of all of the three-jet 
distributions $\chi^2$/NDF = 1.03 (45 degrees of freedom). The 
corresponding result for the four-jet distributions is $\chi^2$/NDF = 1.47 
(63 degrees of freedom) if the $X_A$ distribution is included in the 
comparison, and $\chi^2$/NDF = 1.24 (55 degrees of freedom) if the $X_A$ 
distribution is not included. The result for the combined five-jet 
distributions is $\chi^2$/NDF = 1.21 (63 degrees of freedom). 
The observed distributions are described less well by the HERWIG parton 
shower Monte Carlo predictions, for which the $X_4$, 
$\cos\theta_{3'}$, $\psi_{3'}$, 
and $\cos\theta_{3''}$ distributions have $\chi^2$s significantly 
poorer than those for the corresponding NJETS predictions. Restricting the 
comparison to those distributions predicted by both the NJETS and HERWIG 
calculations (i.e. all distributions except the single-body mass fraction 
distributions) we find the overall $\chi^2$ per degree of freedom  
for the HERWIG comparison of the combined 
three-jet distributions is $\chi^2$/NDF = 1.58 (45 degrees of freedom), 
for the combined four-jet distributions $\chi^2$/NDF = 1.63 (63 degrees 
of freedom), and for the combined five-jet distributions 
$\chi^2$/NDF = 1.52 (63 degrees of freedom).
\end{subsection}

\end{section}

\begin{section} {Conclusions}

The properties of high-mass three-jet, four-jet, and five-jet events 
produced at the Fermilab Tevatron proton-antiproton collider have 
been compared with 
NJETS LO QCD matrix element predictions, HERWIG QCD parton shower 
Monte Carlo predictions, and predictions from a model in which 
events are distributed uniformly over the available multibody 
phase-space. The phase-space model is unable to describe the shapes 
of multijet distributions in regions of parameter space where the QCD 
calculations predict large contributions from initial- and 
final-state gluon radiation. In contrast, the QCD predictions give a 
good first description of the observed multijet distributions, which 
correspond to (4N-4) variables that span the N-body parameter space. 
A more quantitative assessment based on the $\chi^2$ 
per degree of freedom that characterizes the agreement between the observed 
distributions and the QCD predictions shows that NJETS gives a good description 
of all the distributions except perhaps the $X_A$ distribution for 
four-jet events. The NJETS predictions seem to give a better 
description of the observed distributions than the HERWIG predictions. 
This is particularly true of the $X_4$, $\cos\theta_{3'}$, $\psi_{3'}$, 
and $\cos\theta_{3''}$ distributions. 
Finally, we do not see clear evidence for any deviation from the predicted 
multijet distributions that might indicate new phenomena associated 
with the presence of many hard partons in the final state. The general 
agreement between data, NJETS, and HERWIG suggests that $2 \rightarrow 2$ 
scattering plus gluon radiation provides a good first approximation to 
the full LO QCD matrix element for events with three, four, or even five 
jets in the final state. 
\end{section}

\vspace{0.5cm}

{\bf Acknowledgements} 
\vspace{0.2cm}\\
We thank the Fermilab Accelerator Division and the technical and support
staff of our respective institutions. This work was 
supported by the U.S. Department of Energy, the U.S. National Science 
Foundation, the Istituto Nazionale di Fisica Nucleare of Italy, the 
Ministry of Science, Culture and Education of Japan, the Natural Sciences 
and Engineering Research Council of Canada, and the A.P. Sloan Foundation.

\newpage
\vspace*{-10mm}
\renewcommand{\arraystretch}{0.5}
  \baselineskip 0.5cm
\[\begin{array}{lcccc}
\hline
\hline
{\rm Variable}&{\rm NDF}&{\rm NJETS-DATA}&{\rm HERWIG-DATA}&{\rm NJETS-HERWIG}\\
\hline
\spc m_{3J}         & 6& 1.55 & 1.10 & 0.45 \\
\spc X_{3}          & 8& 0.15 & 1.56 & 2.35 \\
\spc X_{4}          & 6& 2.03 & 3.27 & 3.41 \\
\spc \cos\theta_{3} &11& 1.13 & 0.74 & 1.29 \\
\spc \psi_{3}       &14& 0.79 & 1.71 & 2.62 \\
\spc f_{3}          & 9& -    & 2.92 & - \\
\spc f_{4}          & 9& -    & 8.21 & - \\
\spc f_{5}          & 6& -    & 0.30 & - \\
 \hline
\spc m_{4J}         & 6& 0.98 & 1.14 & 0.29 \\
\spc X_{3'}         & 7& 1.37 & 1.00 & 1.61 \\
\spc X_{4'}         & 6& 0.85 & 0.41 & 1.79 \\
\spc \cos\theta_{3'}&15& 1.27 & 2.28 & 1.98 \\
\spc \psi_{3'}      &14& 1.35 & 2.19 & 1.96 \\
\spc X_{A}          & 8& 3.08 & 1.03 & 2.09 \\
\spc \psi'_{AB}     & 7& 1.35 & 1.87 & 1.54 \\
\spc f_{3'}         &11& -    & 3.18 & - \\
\spc f_{4'}         & 8& -    & 3.74 & - \\
\spc f_{5'}         & 8& -    & 1.90 & - \\
\spc f_{A}          &13& -    & 2.20 & - \\
\spc f_{B}          &11& -    & 4.07 & - \\
\hline
\spc m_{5J}          & 8& 0.86 & 1.24 & 1.79 \\
\spc X_{3''}         & 7& 0.98 & 0.80 & 1.28 \\
\spc X_{4''}         & 6& 1.02 & 1.72 & 0.74 \\
\spc \cos\theta_{3''}& 7& 0.61 & 3.19 & 6.20 \\
\spc \psi_{3''}      & 7& 0.68 & 1.85 & 2.11 \\
\spc X_{A'}          & 7& 2.80 & 2.02 & 1.38 \\
\spc \psi''_{A'B'}   & 7& 1.16 & 1.11 & 0.29 \\
\spc X_C             & 7& 1.64 & 0.58 & 1.42 \\
\spc \psi''_{CD}     & 7& 1.09 & 1.27 & 0.17 \\
\spc f_{3''}         &12& -    & 4.11 & - \\
\spc f_{4''}         & 8& -    & 5.66 & - \\
\spc f_{5''}         & 8& -    & 2.82 & - \\
\spc f_C             &10& -    & 1.30 & - \\
\spc f_D             & 7& -    & 5.95 & - \\
\spc f_{A'}          &12& -    & 3.74 & - \\
\spc f_{B'}          &12& -    & 1.57 & - \\
\hline
\hline
\end{array} \]
\renewcommand{\arraystretch}{1.0}
\baselineskip 0.5cm
\vspace*{-\baselineskip}
$\mbox{ }$\\
Table 1: Statistical comparison of agreement between oberserved and 
predicted distributions. The $\chi^2$ per degree of freedom are listed 
for comparisons of the various observed and QCD predicted distributions 
shown in the figures.

\clearpage

\clearpage

\clearpage
\clearpage
%

\clearpage

\clearpage

\clearpage
\begin{figure}
\parbox[t]{6.0in}{
\hspace{-1cm}
\vspace{-1.5in}
\begin{center}
\leavevmode
\epsfysize=6.0in
\epsffile[20 143 575.75 698.75]{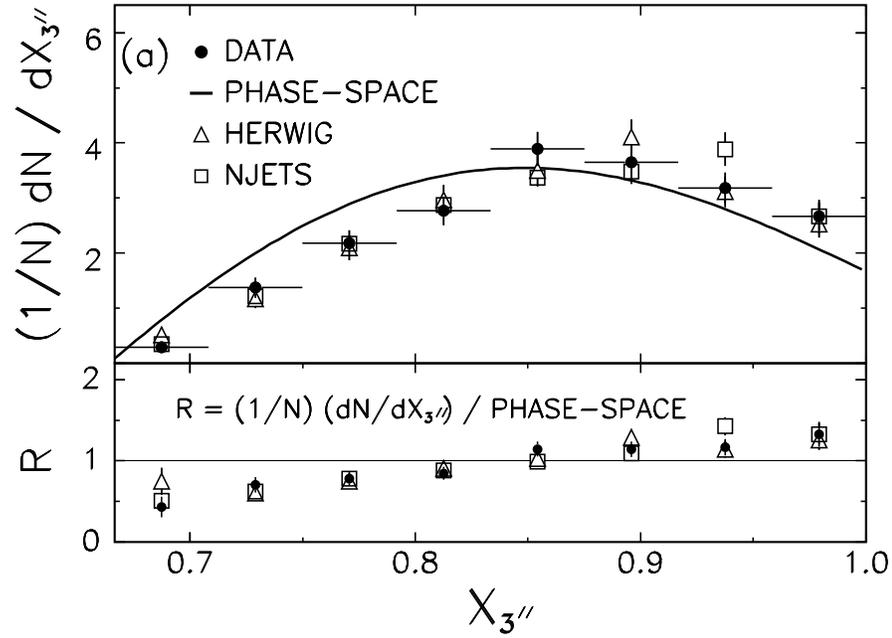}
\end{center}
}
\parbox[b]{6.0in}{
\hspace{-1cm}
\vspace{-2.6in}
\begin{center}
\leavevmode
\epsfysize=6.0in
\epsffile[20 143 575.75 698.75]{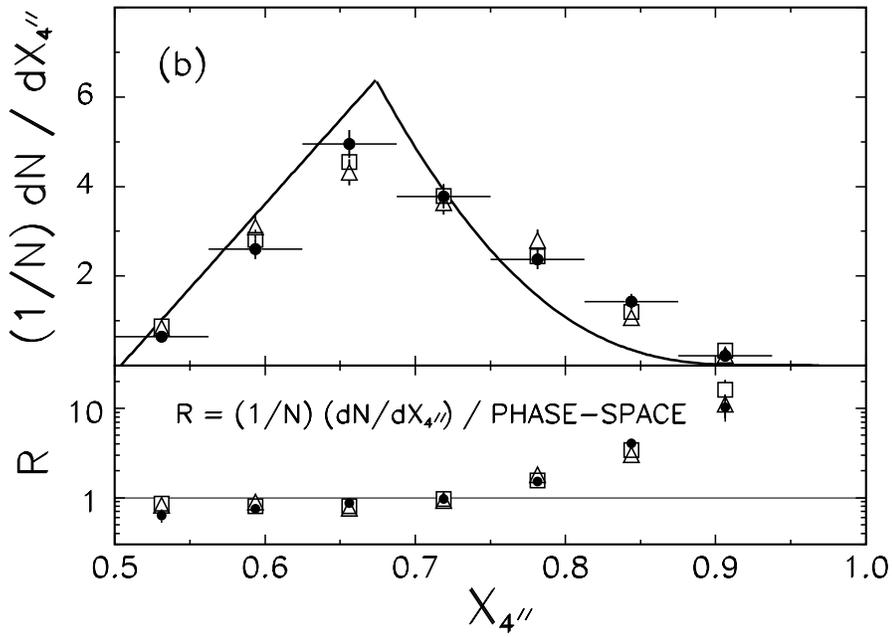}
\end{center}
}
\vspace{-1.5in}
\caption{Dalitz distributions for inclusive five-jet topologies 
that satisfy the requirement $m_{5J}>750$ GeV/$c^2$.
Data (points) are compared with HERWIG predictions (triangles), NJETS 
predictions (squares), and phase-space model predictions (curves) for 
(a) $X_{3''}$, and 
(b) $X_{4''}$.}
\label{fig:5jet_x3_x4}
\end{figure}

\clearpage
\begin{figure}
\parbox[t]{6.0in}{
\hspace{-1cm}
\vspace{-0.5in}
\begin{center}
\leavevmode
\epsfysize=6.0in
\epsffile[20 143 575.75 698.75]{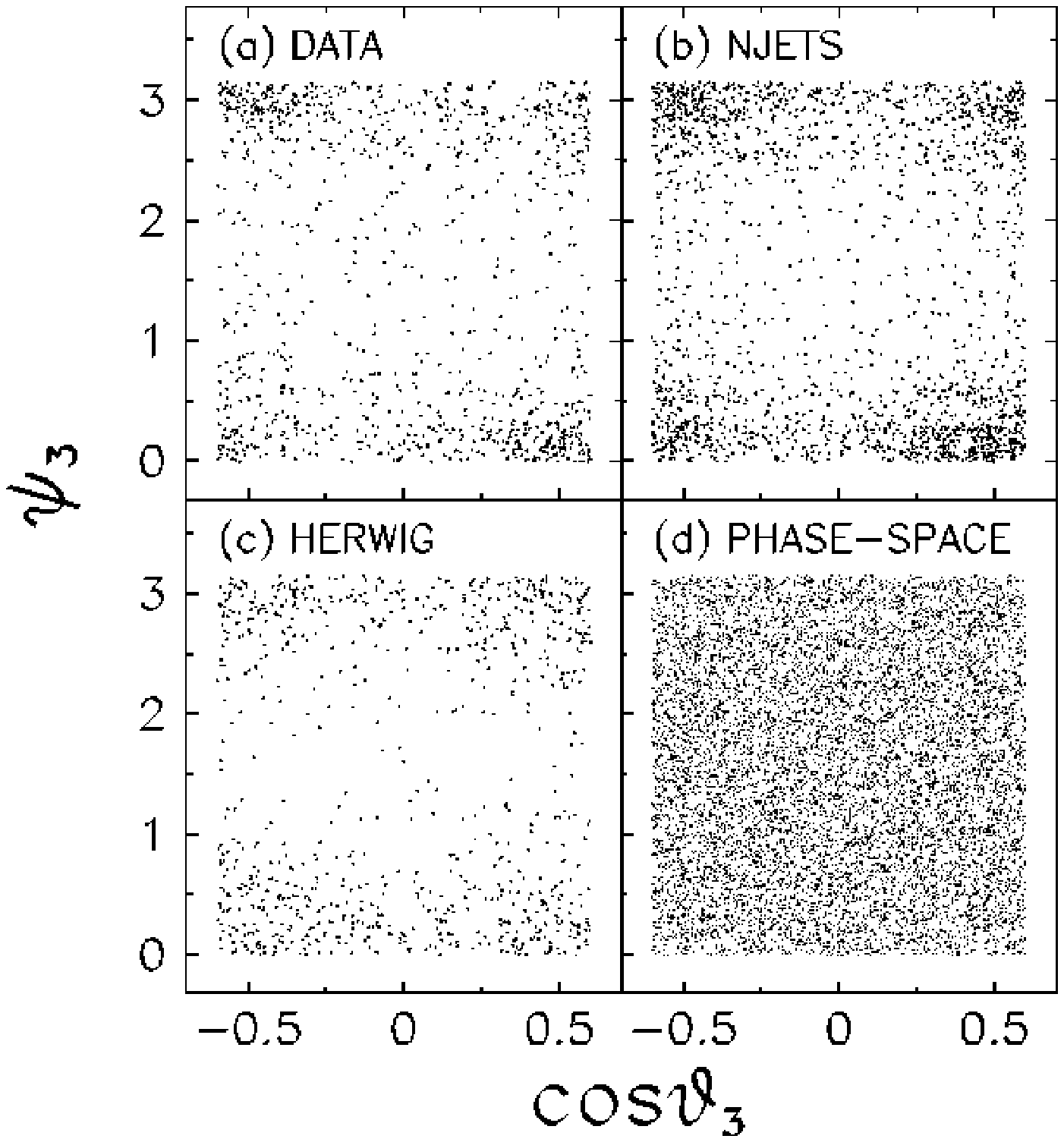}
\end{center}
}
\caption{Inclusive three-jet angular distributions 
for events that satisfy the requirements 
$m_{3J} > 600$ GeV/$c^2$, $X_{3} < 0.9$, and $| \cos\theta_{3}| < 0.6$. 
Event populations in the ($\cos\theta_{3}$,~$\psi_3$)-plane are 
shown for (a) data, (b) NJETS, (c) HERWIG, and (d) phase-space model 
predictions.}
\label{fig:3jet_cos_vs_psi}
\end{figure}

\clearpage
\begin{figure}
\parbox[t]{4.5in}{
\hspace{1.0in}
\vspace{-1.5in}
\begin{center}
\leavevmode
\epsfysize=4.5in
\epsffile[-75 143 575.75 698.75]{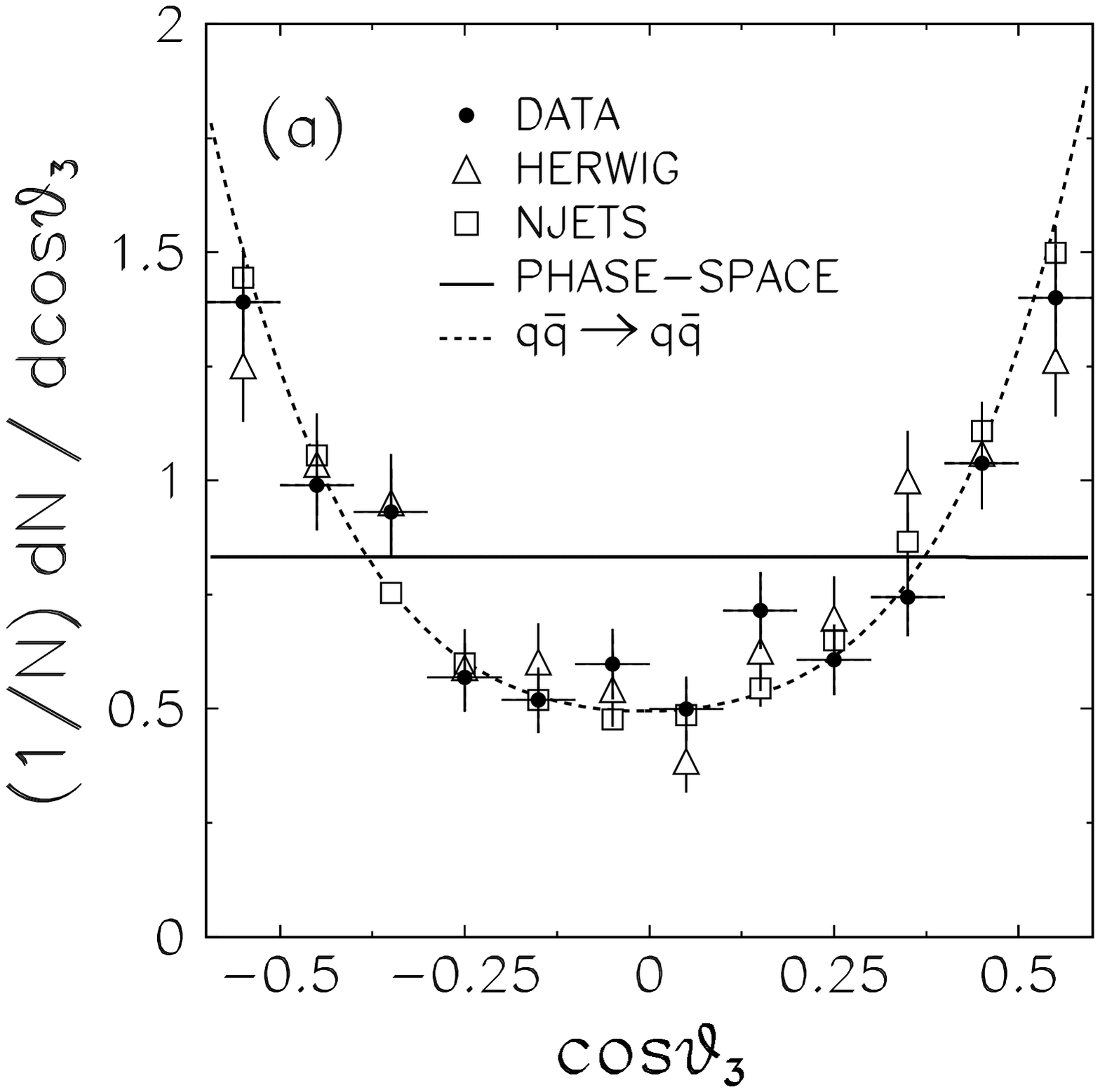}
\end{center}
}
\parbox[b]{4.5in}{
\hspace{1.0 in}
\vspace{-1.35in}
\begin{center}
\leavevmode
\epsfysize=4.5in
\epsffile[-75 143 575.75 698.75]{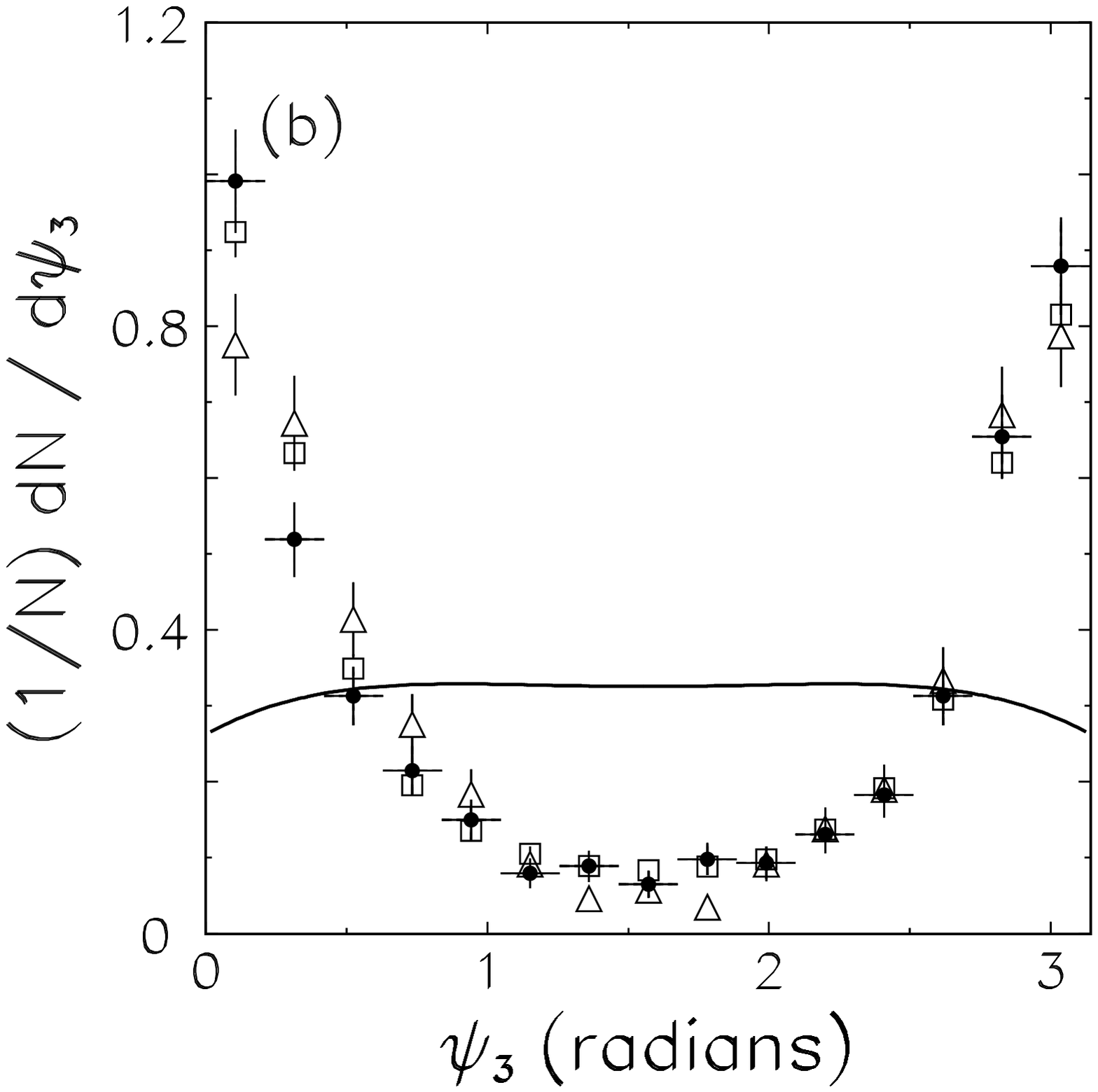}
\end{center}
}
\vspace{-0.2in}
\caption{Inclusive three-jet angular distributions for events that satisfy 
the requirements 
$m_{3J} > 600$ GeV/$c^2$, $X_{3} < 0.9$, and $| \cos\theta_{3}| < 0.6$. 
Data (points) are compared with HERWIG predictions (triangles), NJETS 
predictions (squares), and phase-space model predictions (curves) for 
(a) $\cos\theta_{3}$ and 
(b) $\psi_{3}$.
The broken curve in the $\cos \theta_{3}$ figure is the LO QCD prediction
for $q\overline{q} \rightarrow q\overline{q}$ scattering.}
\label{fig:3jet_cos_psi}
\end{figure}
\clearpage
\begin{figure}
\parbox[t]{6.0in}{
\hspace{-1cm}
\vspace{-0.5in}
\begin{center}
\leavevmode
\epsfysize=6.0in
\epsffile[20 143 575.75 698.75]{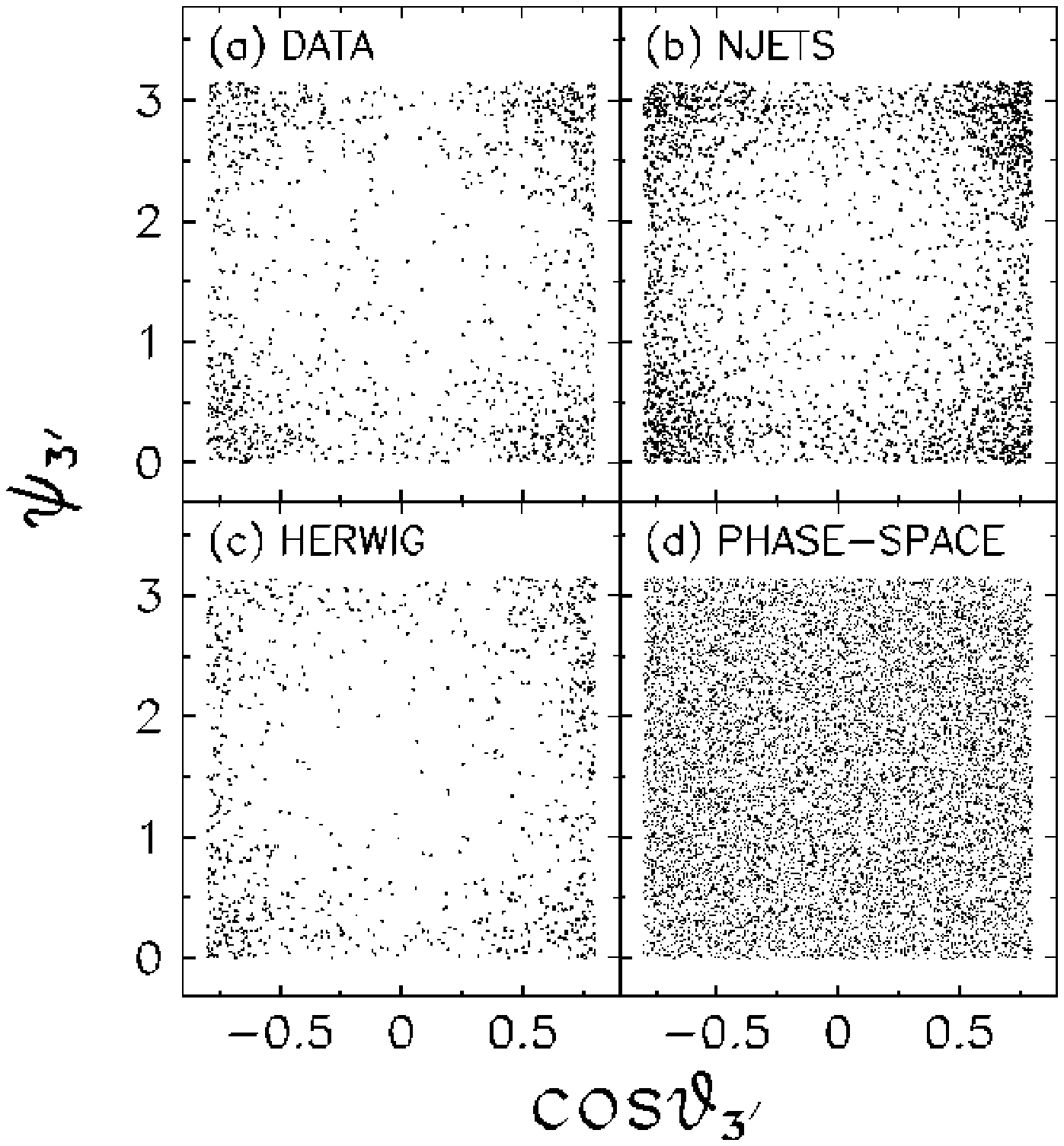}
\end{center}
}
\caption{Inclusive four-jet angular distributions for events 
that satisfy the requirements 
$m_{4J} > 650$ GeV/$c^2$, $X_{3'} < 0.9$, and $| \cos\theta_{3'}| < 0.8$. 
Event populations in the 
($\cos\theta_{3'}$, $\psi_{3'}$)-plane are shown 
for (a) data, (b) NJETS, (c) HERWIG, and (d) phase-space model predictions.} 
\label{fig:4jet_cos_vs_psi}
\end{figure}

\clearpage
\begin{figure}
\parbox[t]{4.5in}{
\hspace{1.0in}
\vspace{-1.5in}
\begin{center}
\leavevmode
\epsfysize=4.5in
\epsffile[-75 143 575.75 698.75]{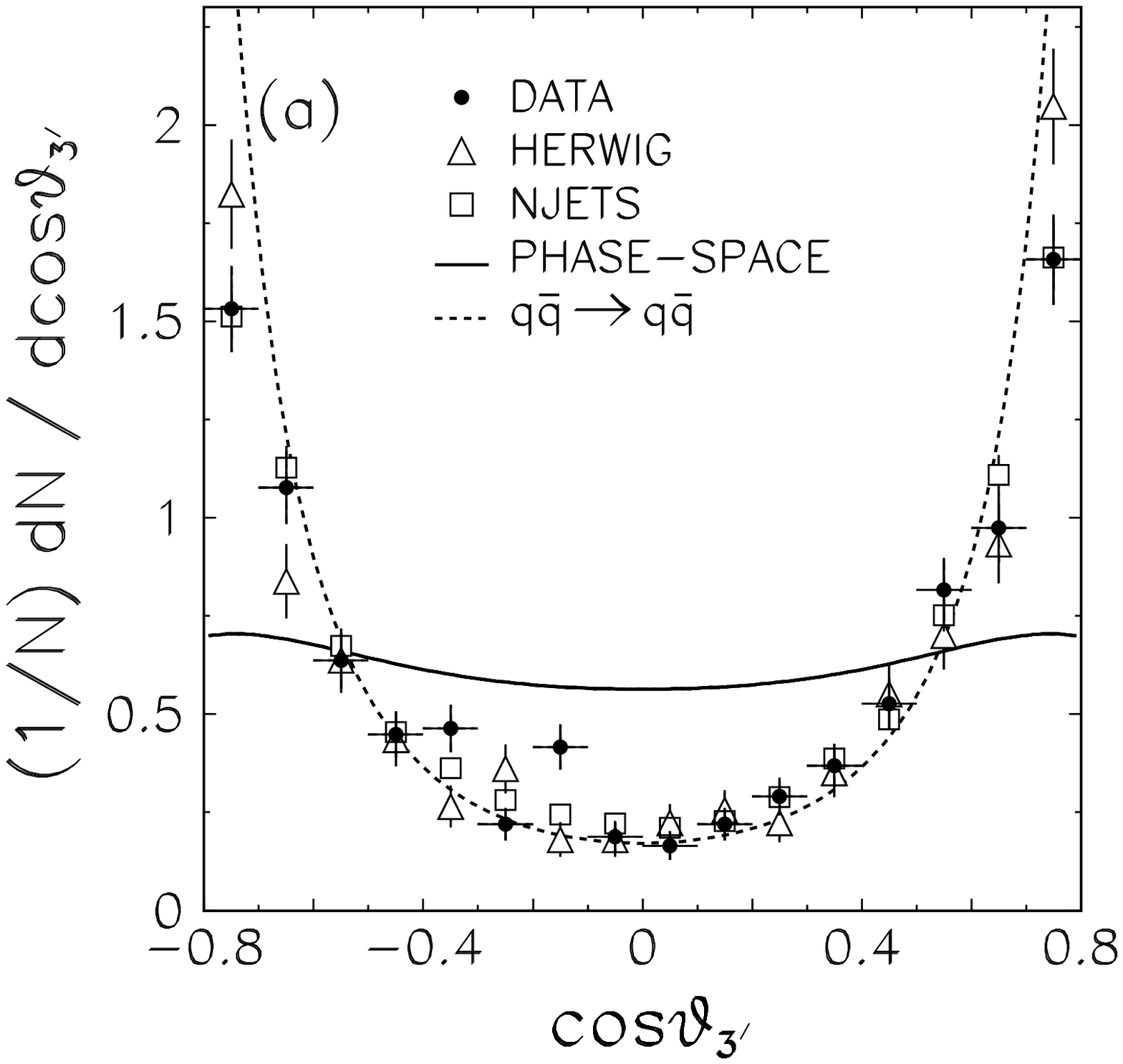}
\end{center}
}
\parbox[b]{4.5in}{
\hspace{1.0 in}
\vspace{-1.35in}
\begin{center}
\leavevmode
\epsfysize=4.5in
\epsffile[-75 143 575.75 698.75]{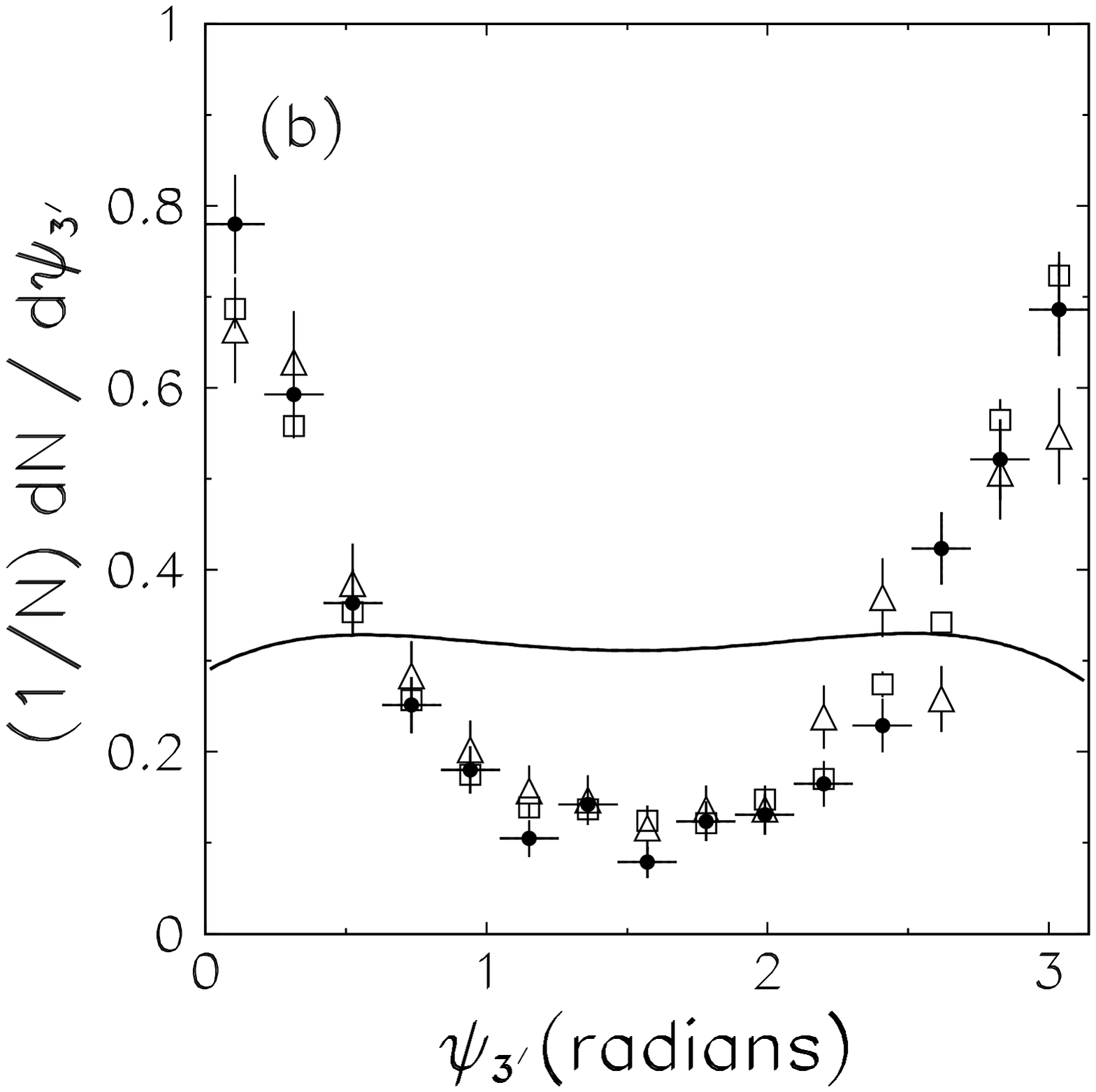}
\end{center}
}
\vspace{-0.2in}
\caption{Inclusive four-jet angular distributions for events 
that satisfy the requirements 
$m_{4J} > 650$ GeV/$c^2$, $X_{3'} < 0.9$, and $| \cos\theta_{3'}| < 0.8$.
Data (points) are compared with HERWIG predictions (triangles), NJETS 
predictions (squares), and phase-space model predictions (curves) for 
(a) $\cos\theta_{3'}$ and 
(b) $\psi_{3'}$.
The broken curve in the $\cos \theta_{3'}$ figure is the LO QCD prediction
for $q\overline{q} \rightarrow q\overline{q}$ scattering.}
\label{fig:4jet_cos_psi}
\end{figure}
\clearpage
\begin{figure}
\parbox[t]{6.0in}{
\hspace{-1cm}
\vspace{-0.5in}
\begin{center}
\leavevmode
\epsfysize=6.0in
\epsffile[20 143 575.75 698.75]{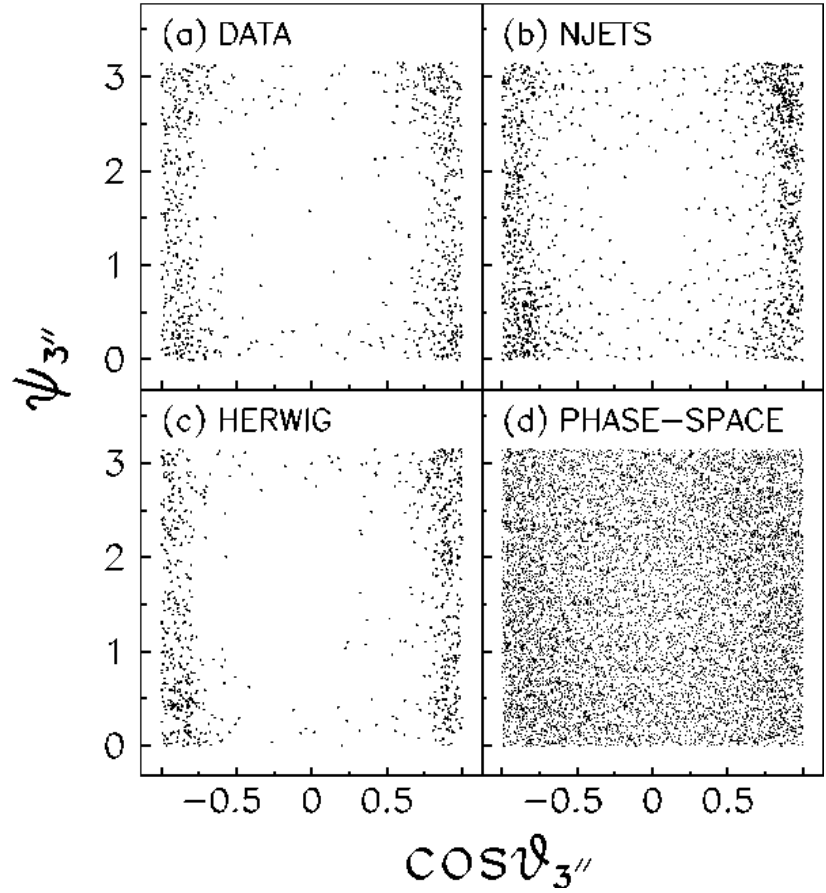}
\end{center}
}
\caption{Inclusive five-jet angular distributions for events 
that satisfy the requirement $m_{5J} > 750$ GeV/$c^2$. 
Event populations in the ($\cos\theta_{3''}$, $\psi_{3''}$)-plane 
are shown 
for (a) data, (b) NJETS, (c) HERWIG, and (d) phase-space model predictions.}
\label{fig:5jet_cos_vs_psi}
\end{figure}

\clearpage
\begin{figure}
\parbox[t]{4.5in}{
\hspace{1.0in}
\vspace{-1.5in}
\begin{center}
\leavevmode
\epsfysize=4.5in
\epsffile[-75 143 575.75 698.75]{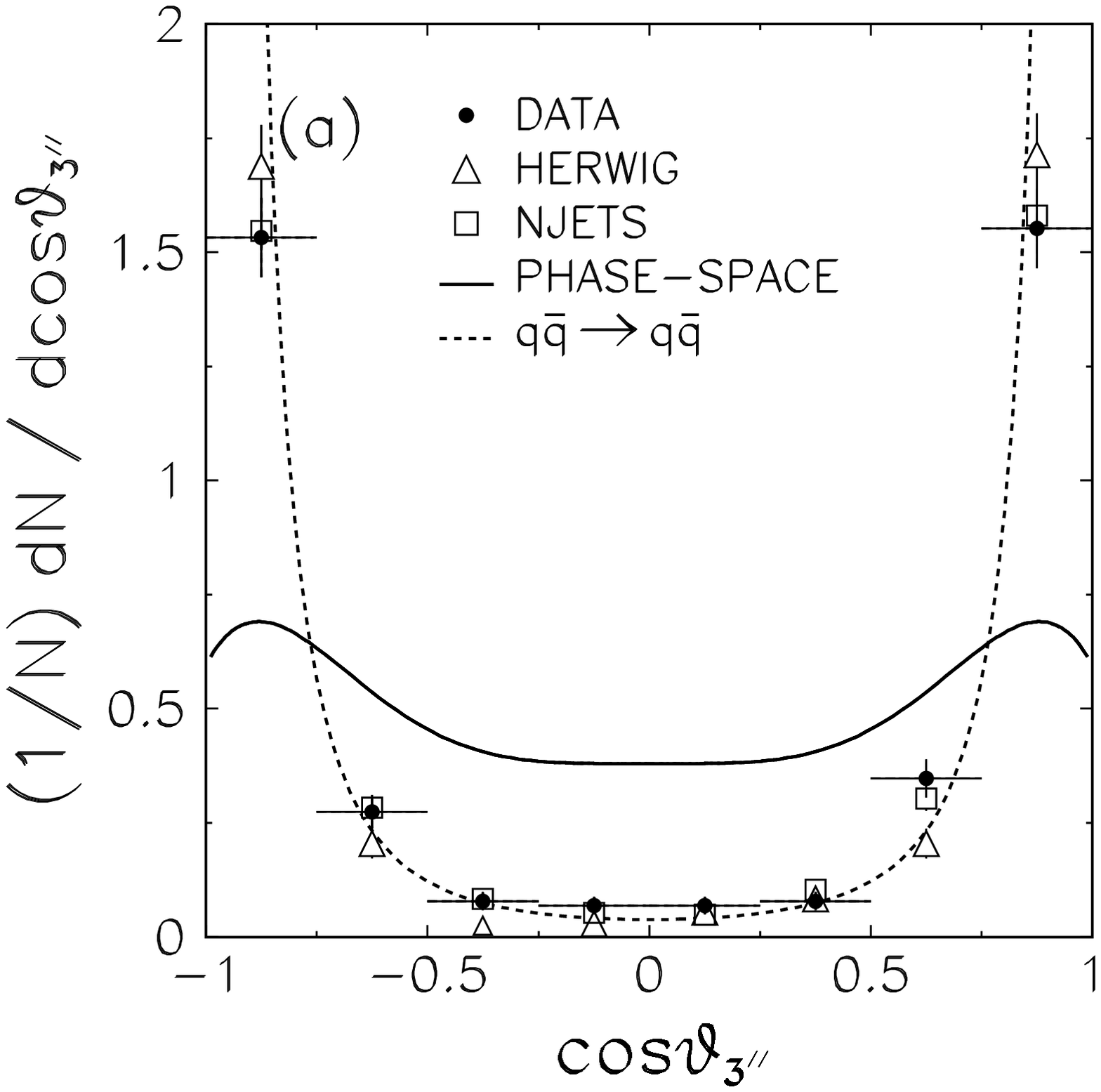}
\end{center}
}
\parbox[b]{4.5in}{
\hspace{1.0 in}
\vspace{-1.35in}
\begin{center}
\leavevmode
\epsfysize=4.5in
\epsffile[-75 143 575.75 698.75]{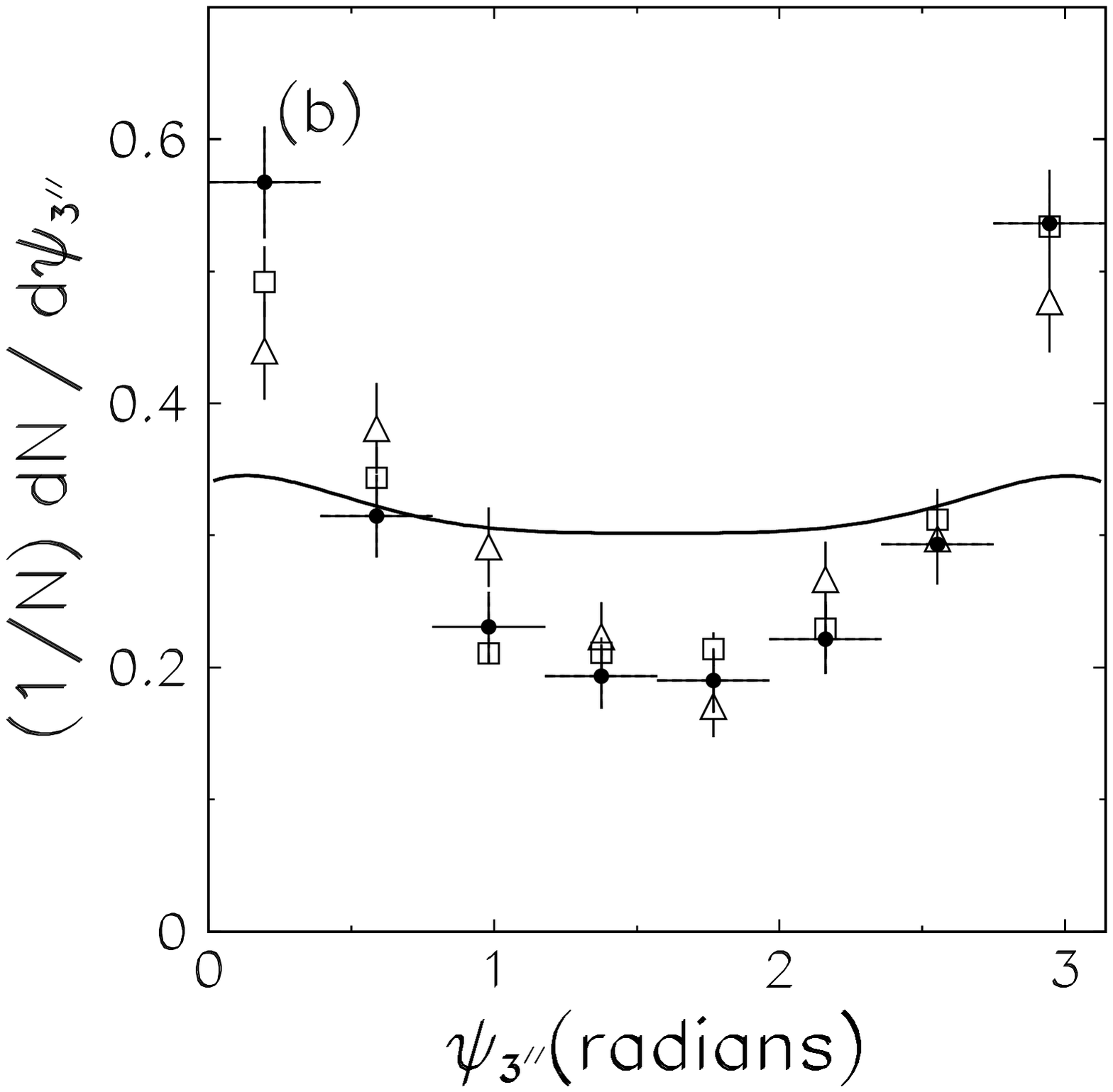}
\end{center}
}
\vspace{-0.2in}
\caption{Inclusive five-jet angular distributions for events 
that satisfy the requirement 
$m_{5J} > 750$ GeV/$c^2$. Data (points) are compared with HERWIG predictions 
(triangles), NJETS predictions (squares), and phase-space model 
  predictions (curves) for (a) $\cos\theta_{3''}$ and
  (b) $\psi_{3''}$.
The broken curve in the $\cos \theta_{3''}$ figure is the LO QCD prediction
for $q\overline{q} \rightarrow q\overline{q}$ scattering.}
\label{fig:5jet_cos_psi}
\end{figure}

%
\clearpage
\begin{figure}
\parbox[t]{6.0in}{
\hspace{-1cm}
\vspace{-0.5in}
\begin{center}
\leavevmode
\epsfysize=6.0in
\epsffile[20 143 575.75 698.75]{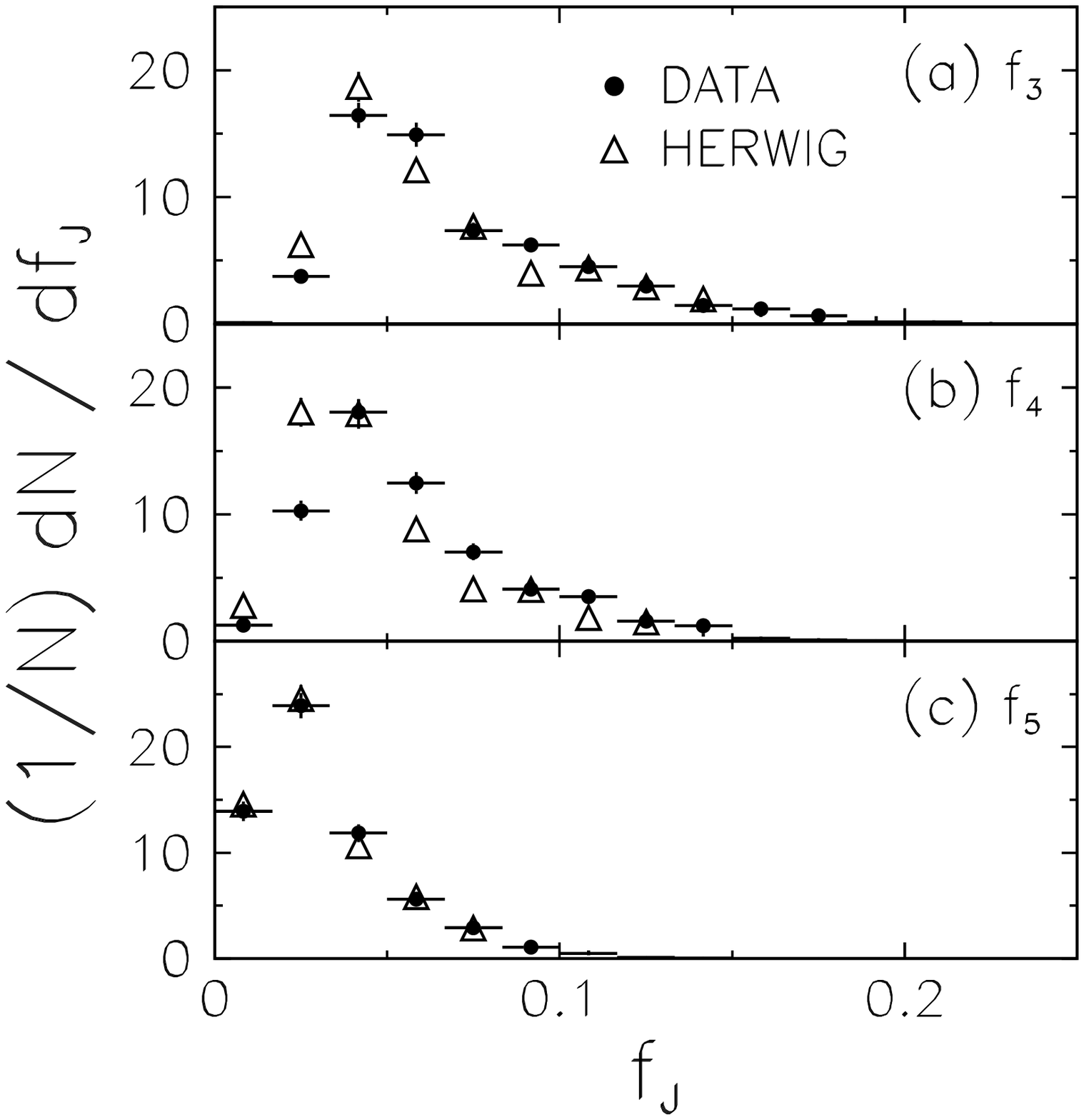}
\end{center}
}
\caption{Single-jet mass fraction distributions for inclusive 
three-jet events. Data (points) compared with
HERWIG predictions (triangles), shown for (a) the highest energy 
jet in the three-jet rest-frame, (b) the second-to-highest energy 
jet, and (c) the third-to-highest energy jet.}
\label{fig:3jet_f345}
\end{figure}
%
%
\clearpage
\begin{figure}
\parbox[t]{6.0in}{
\hspace{-1cm}
\vspace{-0.5in}
\begin{center}
\leavevmode
\epsfysize=6.0in
\epsffile[20 143 575.75 698.75]{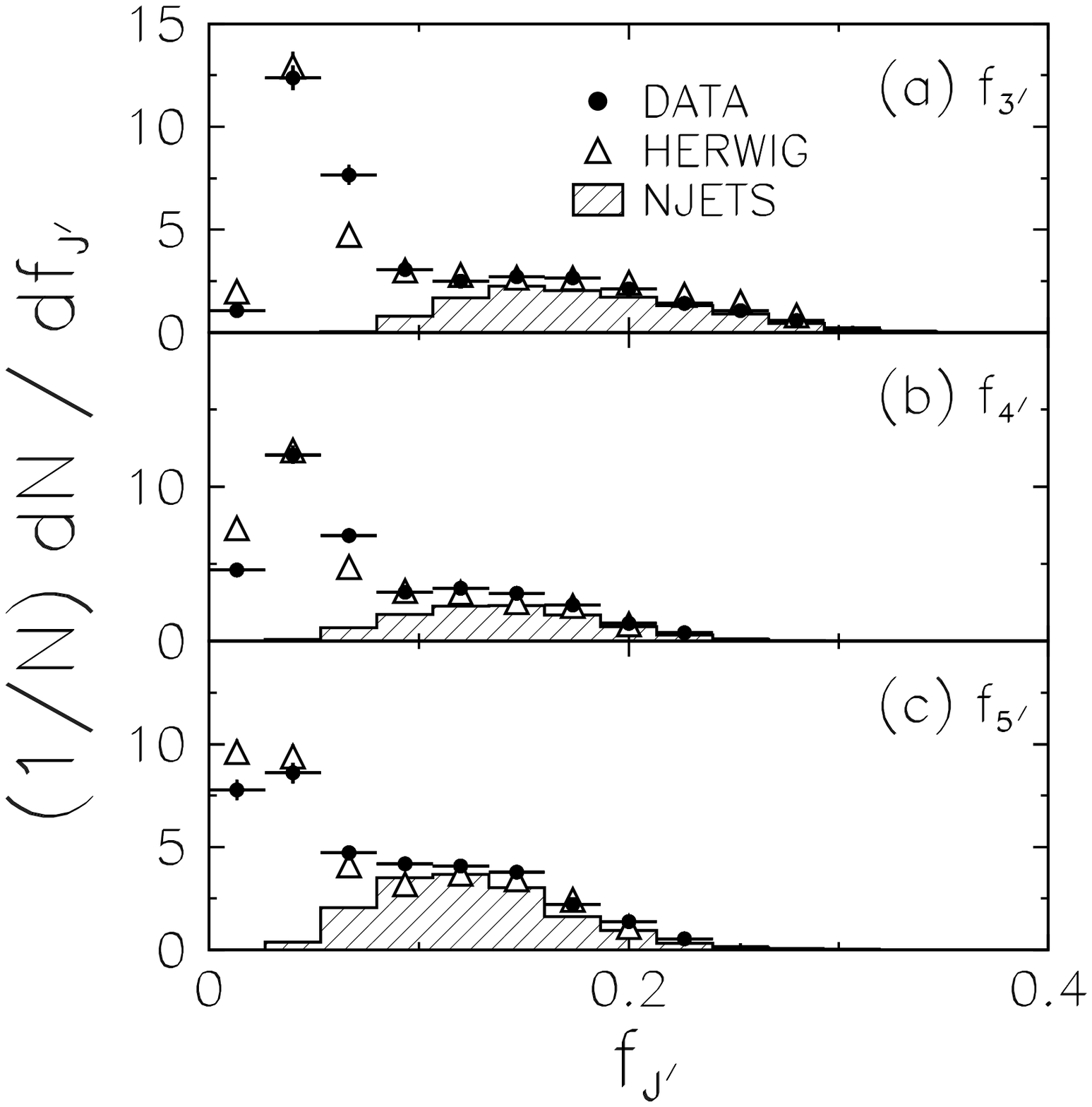}
\end{center}
}
\caption{Single-body mass fraction distributions for inclusive 
four-jet events. Data (points) compared with
HERWIG predictions (triangles), and NJETS predictions (histograms), 
shown for (a) the highest energy body 
in the three-body rest-frame, (b) the second-to-highest energy body, 
and (c) the third-to-highest energy body.}
\label{fig:4jet_f345}
\end{figure}
%
%
\clearpage
\begin{figure}
\parbox[t]{6.0in}{
\hspace{-1cm}
\vspace{-0.5in}
\begin{center}
\leavevmode
\epsfysize=6.0in
\epsffile[20 143 575.75 698.75]{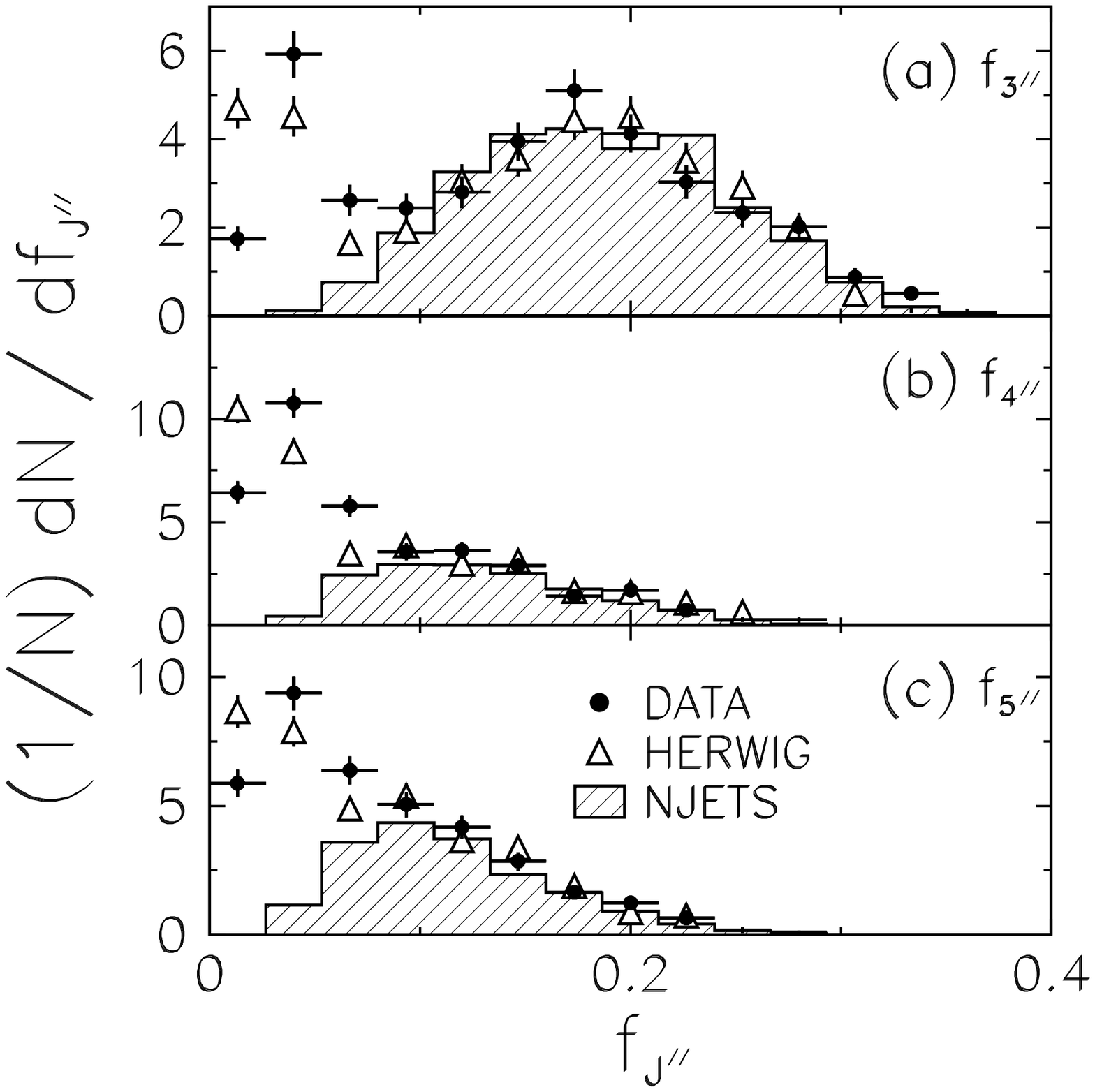}
\end{center}
}
\caption{Single-body mass fraction distributions for inclusive 
five-jet events. Data (points) compared with
HERWIG predictions (triangles), and NJETS predictions (histograms), 
shown for (a) the highest energy body 
in the three-body rest-frame, (b) the second-to-highest energy body, 
and (c) the third-to-highest energy body.}
\label{fig:5jet_f345}
\end{figure}
%
%
\begin{figure}
\parbox[t]{6.0in}{
\hspace{-1cm}
\vspace{-0.5in}
\begin{center}
\leavevmode
\epsfysize=6.0in
\epsffile[20 143 575.75 698.75]{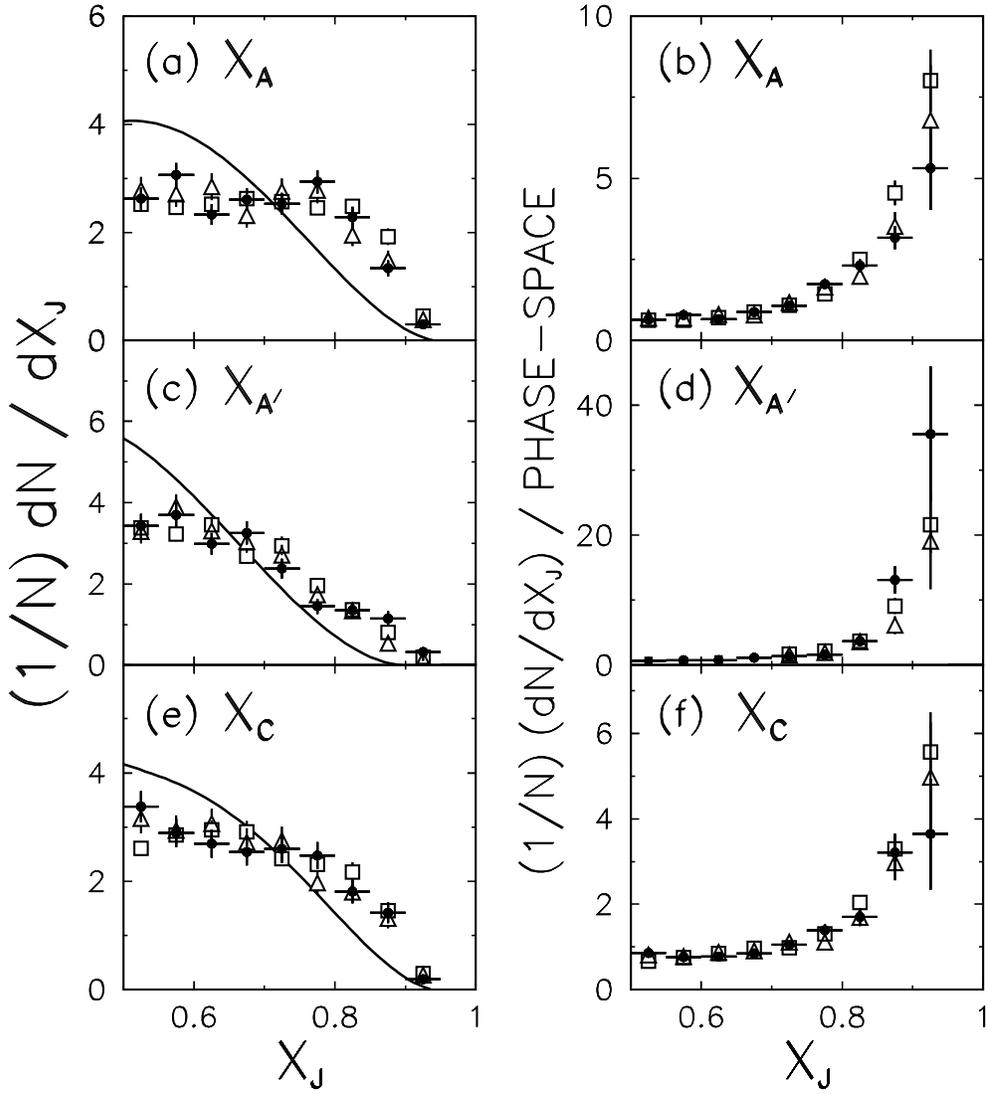}
\end{center}
}
\caption{The two-body energy sharing distributions for inclusive 
four-jet and five-jet events.
Data (points) are compared with HERWIG predictions (triangles), 
NJETS predictions (squares), and phase-space predictions (curves)
for 
(a) $X_{A}$, 
(b) $X_{A}$ after dividing by the phase-space model predictions, 
(c) $X_{A'}$, 
(d) $X_{A'}$ after dividing by the phase-space model predictions, 
(e) $X_{C} $, and 
(f) $X_{C} $ after dividing by the phase-space model predictions.}
\label{fig:45jet_xa_xc}
\end{figure}

%
\begin{figure}
\parbox[t]{6.0in}{
\hspace{-1cm}
\vspace{-0.5in}
\begin{center}
\leavevmode
\epsfysize=6.0in
\epsffile[20 143 575.75 698.75]{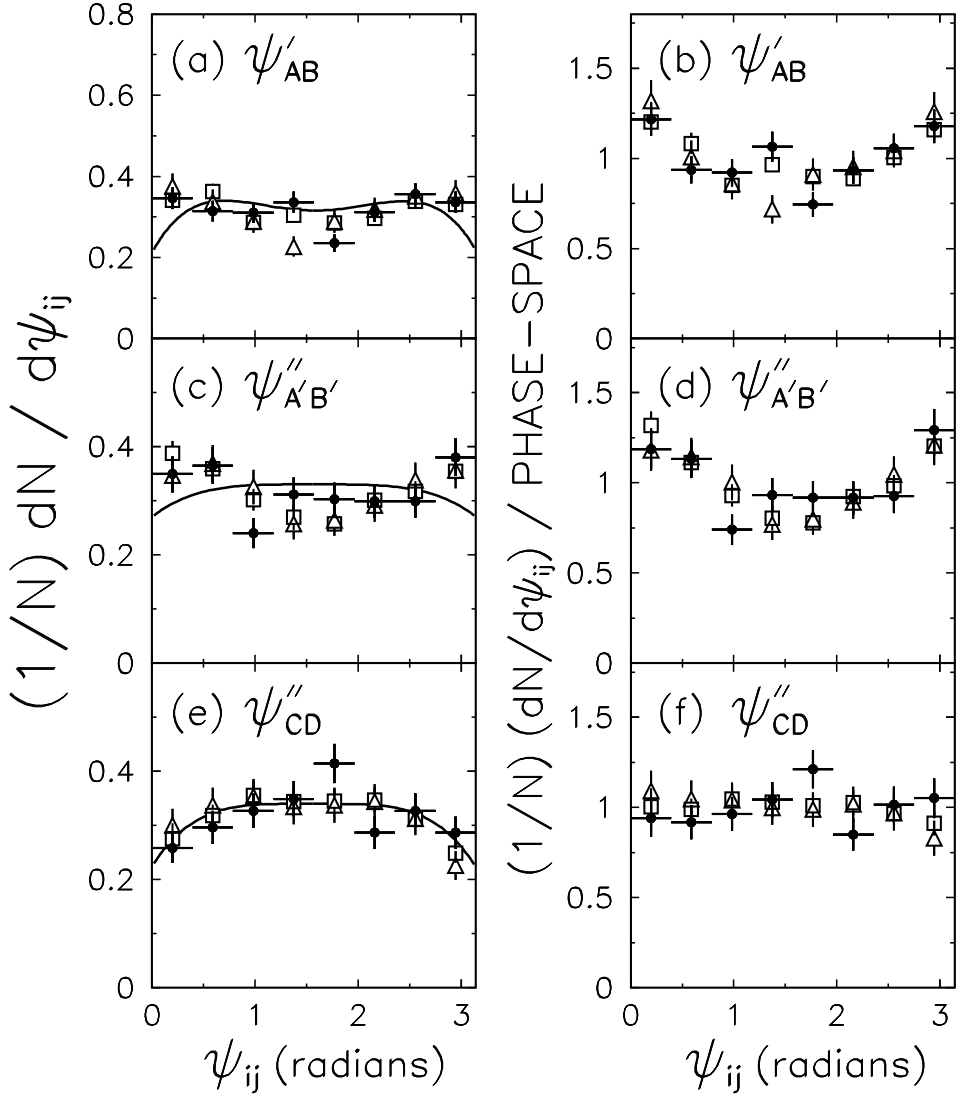}
\end{center}
}
\vspace{-0.2in}
\caption{Two-body angular distributions for inclusive 
four-jet and five-jet events.
Data (points) are compared with HERWIG predictions (triangles), 
NJETS predictions (squares), and phase-space predictions (curves)
for 
(a) $\psi'_{AB}$, 
(b) $\psi'_{AB}$ after dividing by the phase-space model predictions,
(c) $\psi''_{A'B'}$, 
(d) $\psi''_{A'B'}$ after dividing by the phase-space model predictions,
(e) $\psi''_{CD}$, and 
(f) $\psi''_{CD}$ after dividing by the phase-space model predictions.} 
\label{fig:45jet_pab_pcd}
\end{figure}
%

%
\begin{figure}
\parbox[t]{7.0in}{
\hspace{-1cm}
\vspace{-0.5in}
\begin{center}
\leavevmode
\epsfysize=7.0in
\epsffile[20 143 575.75 698.75]{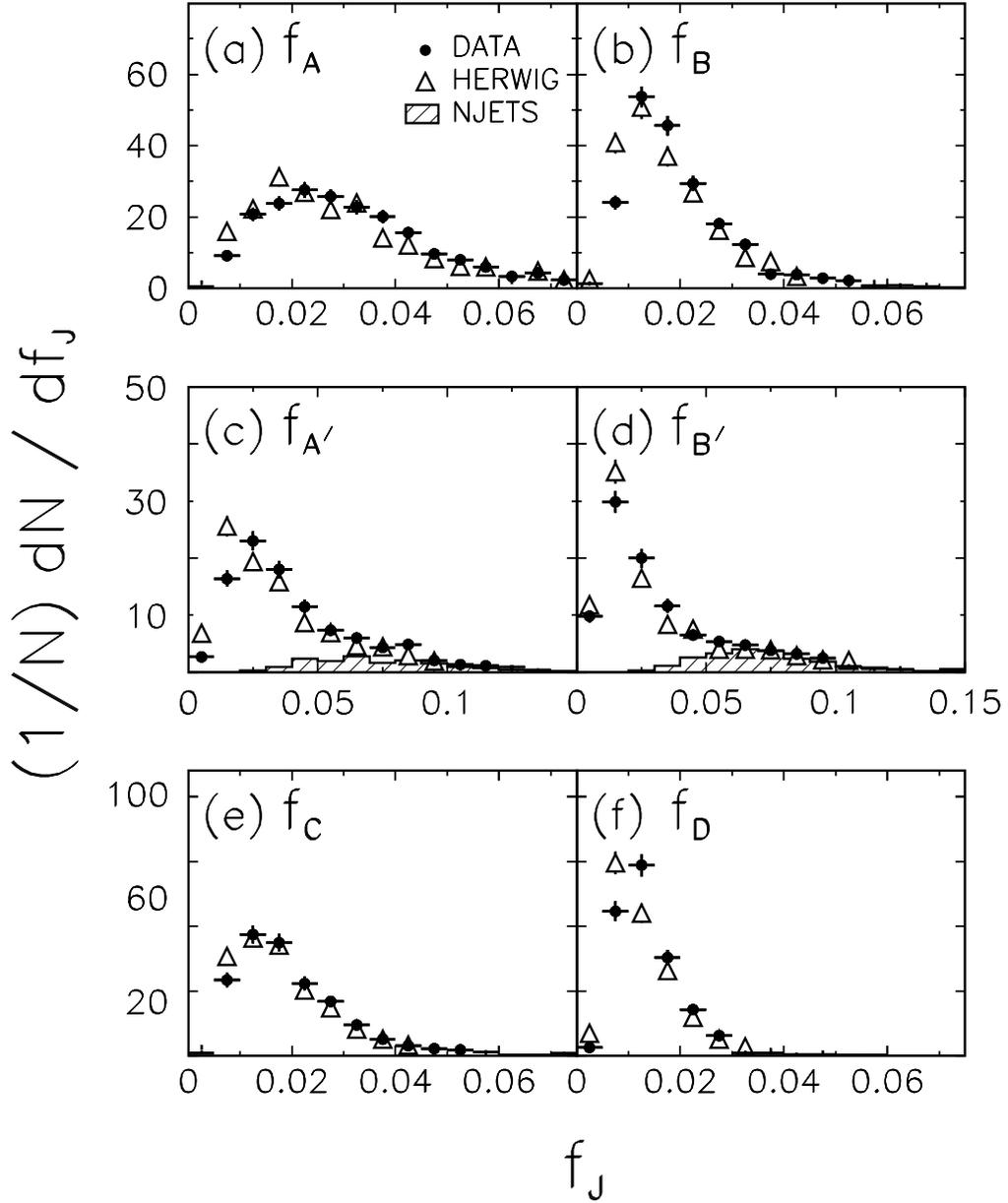}
\end{center}
}
\vspace{-0.2in}
\caption{Single-body mass fraction distributions for two-body systems
in inclusive four-jet and five-jet events.
Data (points) are compared with HERWIG predictions (triangles), 
and NJETS predictions (histograms) for 
(a) $f_{A}$, 
(b) $f_{B}$, 
(c) $f_{A'}$, 
(d) $f_{B'}$, 
(e) $f_{C}$, and 
(f) $f_{D}$.}
\label{fig:45jet_fabcd}
\end{figure}

\end{document}